# Identification of Galaxy–Galaxy Strong Lens Candidates in the DECam Local Volume Exploration Survey Using Machine Learning

E. A. Zaborowski[1,2,3], A. Drlica-Wagner[3,4,5], F. Ashmead[4], J. F. Wu[6,7], R. Morgan[8], C. R. Bom[9], A. J. Shajib[3,5,70], S. Birrer[10,11], W. Cerny[3,5,12], E. J. Buckley-Geer[3,4], B. Mutlu-Pakdil[13], P. S. Ferguson[14], K. Glazebrook[15], S. J. Gonzalez Lozano[8], Y. Gordon[8], M. Martinez[14], V. Manwadkar[5], J. O'Donnell[16], J. Poh[3,5], A. Riley[17,18,19], J. D. Sakowska[20], L. Santana-Silva[21], B. X. Santiago[22,23], D. Sluse[24], C. Y. Tan[3,25], E. J. Tollerud[6], A. Verma[26], J. A. Carballo-Bello[27], Y. Choi[28], D. J. James[29], N. Kuropatkin[4], C. E. Martínez-Vázquez[30], D. L. Nidever[31], J. L. Nilo Castellon[32], N. E. D. Noël[33], K. A. G. Olsen[34], A. B. Pace[35], S. Mau[10,36], B. Yanny[4], A. Zenteno[37], T. M. C. Abbott[37], M. Aguena[23], O. Alves[38], F. Andrade-Oliveira[38], S. Bocquet[39], D. Brooks[40], D. L. Burke[10,11], A. Carnero Rosell[23,41,42], M. Carrasco Kind[43,44], J. Carretero[45], F. J. Castander[46,47], C. J. Conselice[48], M. Costanzi[49,50,51], M. E. S. Pereira[52], S. De Vicente[53], S. Desai[54], J. P. Dietrich[39], P. Doel[40], S. Everett[55], I. Ferrero[56], B. Flaugher[4], D. Friedel[43], J. Frieman[3,4,5], J. García-Bellido[57], D. Gruen[39], R. A. Gruendl[43,44], G. Gutierrez[4], S. R. Hinton[58], D. L. Hollowood[16], K. Honscheid[1,2], K. Kuehn[59,60], H. Lin[4], J. L. Marshall[18,19], P. Melchior[61], J. Mena-Fernández[53], F. Menanteau[43,44], R. Miquel[45,62], A. Palmese[63], F. Paz-Chinchón[43,64], A. Pieres[23,65], A. A. Plazas Malagón[61], J. Prat[3,5], M. Rodriguez-Monroy[53], A. K. Romer[66], E. Sanchez[53], V. Scarpine[4], I. Sevilla-Noarbe[53], M. Smith[67], E. Suchyta[68], C. To[2], and N. Weaverdyck[38,69]

(DELVE & DES Collaborations)

[1] Department of Physics, The Ohio State University, Columbus, OH 43210, USA; zaborowski.11@osu.edu
[2] Center for Cosmology and Astro-Particle Physics, The Ohio State University, Columbus, OH 43210, USA
[3] Kavli Institute for Cosmological Physics, University of Chicago, Chicago, IL 60637, USA
[4] Fermi National Accelerator Laboratory, P.O. Box 500, Batavia, IL 60510, USA
[5] Department of Astronomy and Astrophysics, University of Chicago, Chicago, IL 60637, USA
[6] Space Telescope Science Institute, 3700 San Martin Drive, Baltimore, MD 21218, USA
[7] Department of Physics & Astronomy, Johns Hopkins University, Baltimore, MD 21218, USA
[8] Physics Department, 2320 Chamberlin Hall, University of Wisconsin-Madison, 1150 University Avenue Madison, WI 53706-1390, USA
[9] Centro Brasileiro de Pesquisas Físicas, Rua Dr. Xavier Sigaud 150, 22290-180 Rio de Janeiro, RJ, Brazil
[10] Kavli Institute for Particle Astrophysics & Cosmology, P.O. Box 2450, Stanford University, Stanford, CA 94305, USA
[11] SLAC National Accelerator Laboratory, Menlo Park, CA 94025, USA
[12] Department of Astronomy, Yale University, New Haven, CT 06520, USA
[13] Department of Physics and Astronomy, Dartmouth College, Hanover, NH 03755, USA
[14] Department of Physics, University of Wisconsin-Madison, Madison, WI 53706, USA
[15] Centre for Astrophysics & Supercomputing, Swinburne University of Technology, VIC 3122, Australia
[16] Santa Cruz Institute for Particle Physics, Santa Cruz, CA 95064, USA
[17] Institute for Computational Cosmology, Department of Physics, Durham University, South Road, Durham, DH1 3LE, UK
[18] George P. and Cynthia Woods Mitchell Institute for Fundamental Physics and Astronomy, Texas A&M University, College Station, TX 77843, USA
[19] Department of Physics and Astronomy, Texas A&M University, College Station, TX 77843, USA
[20] Department of Physics, University of Surrey, Guildford, GU2 7XH, UK
[21] NAT-Universidade Cruzeiro do Sul/Universidade Cidade de São Paulo, Rua Galvão Bueno, 868, 01506-000, São Paulo, SP, Brazil
[22] Instituto de Física, UFRGS, Caixa Postal 15051, Porto Alegre, RS—91501-970, Brazil
[23] Laboratório Interinstitucional de e-Astronomia—LIneA, Rua Gal. José Cristino 77, Rio de Janeiro, RJ—20921-400, Brazil
[24] STAR Institute, Quartier Agora—Allée du six Aout, 19c B-4000 Liége, Belgium
[25] Department of Physics, University of Chicago, Chicago, IL 60637, USA
[26] Sub-department of Astrophysics, University of Oxford, Denys Wilkinson Building, Oxford, OX1 3RH, UK
[27] Instituto de Alta Investigación, Sede Esmeralda, Universidad de Tarapacá, Av. Luis Emilio Recabarren 2477, Iquique, Chile
[28] Department of Astronomy, University of California, Berkeley, CA 94720, USA
[29] ASTRAVEO LLC, P.O. Box 1668, MA 01931, USA
[30] Gemini Observatory/NSF's NOIRLab, 670 N. A'ohoku Place, Hilo, HI 96720, USA
[31] Department of Physics, Montana State University, P.O. Box 173840, Bozeman, MT 59717-3840, USA
[32] Dirección Investigación y Desarrollo, Universidad de La Serena, Avenida Juan Cisternas 1200, La Serena, Chile
[33] Physics Department, University of Surrey, Guildford, GU2 7XH, UK
[34] NSF's National Optical Infrared Astronomy Research Laboratory, 950 N. Cherry Avenue, Tucson, AZ 85719, USA
[35] McWilliams Center for Cosmology, Carnegie Mellon University, 5000 Forbes Avenue, Pittsburgh, PA 15213, USA
[36] Department of Physics, Stanford University, 382 Via Pueblo Mall, Stanford, CA 94305, USA
[37] Cerro Tololo Inter-American Observatory, NSF's National Optical-Infrared Astronomy Research Laboratory, Casilla 603, La Serena, Chile
[38] Department of Physics, University of Michigan, Ann Arbor, MI 48109, USA
[39] University Observatory, Faculty of Physics, Ludwig-Maximilians-Universität, Scheinerstr. 1, D-81679 Munich, Germany
[40] Department of Physics & Astronomy, University College London, Gower Street, London, WC1E 6BT, UK
[41] Instituto de Astrofisica de Canarias, E-38205 La Laguna, Tenerife, Spain
[42] Universidad de La Laguna, Dpto. Astrofísica, E-38206 La Laguna, Tenerife, Spain
[43] Center for Astrophysical Surveys, National Center for Supercomputing Applications, 1205 West Clark Street, Urbana, IL 61801, USA
[44] Department of Astronomy, University of Illinois at Urbana-Champaign, 1002 W. Green Street, Urbana, IL 61801, USA
[45] Institut de Física d'Altes Energies (IFAE), The Barcelona Institute of Science and Technology, Campus UAB, E-08193 Bellaterra (Barcelona) Spain
[46] Institute of Space Sciences (ICE, CSIC), Campus UAB, Carrer de Can Magrans, s/n, E-08193 Bellaterra (Barcelona), Spain
[47] Institut d'Estudis Espacials de Catalunya (IEEC), E-08034 Barcelona, Spain
[48] Jodrell Bank Centre for Astrophysics, University of Manchester, Oxford Road, Manchester, M13 9PY, UK
[49] Astronomy Unit, Department of Physics, University of Trieste, via Tiepolo 11, I-34131 Trieste, Italy

[50] INAF-Osservatorio Astronomico di Trieste, via G.B. Tiepolo 11, I-34143 Trieste, Italy
[51] Institute for Fundamental Physics of the Universe, Via Beirut 2, I-34014 Trieste, Italy
[52] Hamburger Sternwarte, Universität Hamburg, Gojenbergsweg 112, D-21029 Hamburg, Germany
[53] Centro de Investigaciones Energéticas, Medioambientales y Tecnológicas (CIEMAT), Madrid, Spain
[54] Department of Physics, IIT Hyderabad, Kandi, Telangana 502285, India
[55] Jet Propulsion Laboratory, California Institute of Technology, 4800 Oak Grove Drive, Pasadena, CA 91109, USA
[56] Institute of Theoretical Astrophysics, University of Oslo, P.O. Box 1029 Blindern, NO-0315 Oslo, Norway
[57] Instituto de Fisica Teorica UAM/CSIC, Universidad Autonoma de Madrid, E-28049 Madrid, Spain
[58] School of Mathematics and Physics, University of Queensland, Brisbane, QLD 4072, Australia
[59] Australian Astronomical Optics, Macquarie University, North Ryde, NSW 2113, Australia
[60] Lowell Observatory, 1400 Mars Hill Rd, Flagstaff, AZ 86001, USA
[61] Department of Astrophysical Sciences, Princeton University, Peyton Hall, Princeton, NJ 08544, USA
[62] Institució Catalana de Recerca i Estudis Avançats, E-08010 Barcelona, Spain
[63] Department of Astronomy, University of California, Berkeley, 501 Campbell Hall, Berkeley, CA 94720, USA
[64] Institute of Astronomy, University of Cambridge, Madingley Road, Cambridge, CB3 0HA, UK
[65] Observatório Nacional, Rua Gal. José Cristino 77, Rio de Janeiro, RJ—20921-400, Brazil
[66] Department of Physics and Astronomy, Pevensey Building, University of Sussex, Brighton, BN1 9QH, UK
[67] School of Physics and Astronomy, University of Southampton, Southampton, SO17 1BJ, UK
[68] Computer Science and Mathematics Division, Oak Ridge National Laboratory, Oak Ridge, TN 37831, USA
[69] Lawrence Berkeley National Laboratory, 1 Cyclotron Road, Berkeley, CA 94720, USA
*Received 2022 November 11; revised 2023 July 1; accepted 2023 July 3; published 2023 August 23*

## Abstract

We perform a search for galaxy–galaxy strong lens systems using a convolutional neural network (CNN) applied to imaging data from the first public data release of the DECam Local Volume Exploration Survey, which contains ∼520 million astronomical sources covering ∼4000 deg$^2$ of the southern sky to a 5$\sigma$ point–source depth of $g = 24.3$, $r = 23.9$, $i = 23.3$, and $z = 22.8$ mag. Following the methodology of similar searches using Dark Energy Camera data, we apply color and magnitude cuts to select a catalog of ∼11 million extended astronomical sources. After scoring with our CNN, the highest-scoring 50,000 images were visually inspected and assigned a score on a scale from 0 (not a lens) to 3 (very probable lens). We present a list of 581 strong lens candidates, 562 of which are previously unreported. We categorize our candidates using their human-assigned scores, resulting in 55 Grade A candidates, 149 Grade B candidates, and 377 Grade C candidates. We additionally highlight eight potential quadruply lensed quasars from this sample. Due to the location of our search footprint in the northern Galactic cap ($b > 10$ deg) and southern celestial hemisphere (decl. < 0 deg), our candidate list has little overlap with other existing ground-based searches. Where our search footprint does overlap with other searches, we find a significant number of high-quality candidates that were previously unidentified, indicating a degree of orthogonality in our methodology. We report properties of our candidates including apparent magnitude and Einstein radius estimated from the image separation.

*Unified Astronomy Thesaurus concepts:* Strong gravitational lensing (1643)

*Supporting material:* machine-readable table

## 1. Introduction

Strong gravitational lensing results from the general-relativistic deflection of light caused by the inhomogeneous distribution of matter along the line of sight. Since the first observations of a doubly imaged lensed quasar (Walsh et al. 1979) and giant arcs of lensed galaxies near the center of galaxy clusters (Lynds & Petrosian 1986; Paczynski 1987; Soucail et al. 1987, 1988), strong lensing has grown into an important tool for constraining astrophysics and cosmology (see Treu 2010, for a review).

Strong lens systems provide a wealth of astrophysical information that is difficult or unfeasible to obtain with traditional analysis. The magnifying effect of strong lensing can enable detailed study of the internal morphologies of background source galaxies down to scales of tens of parsecs for galaxies that are too distant or faint to study under normal circumstances (e.g., Bayliss et al. 2014; Livermore et al. 2015; Johnson et al. 2017; Cornachione et al. 2018; Ritondale et al. 2019a; Rivera-Thorsen et al. 2019; Ivison et al. 2020; Florian et al. 2021; Khullar et al. 2021). Strong lensing can be used to measure the distribution of both dark and luminous matter in galaxies, which provides essential information for modeling galaxy formation, constraining the stellar initial mass function, and understanding the baryonic processes that drive galaxy evolution (e.g., Treu & Koopmans 2002; Czoske et al. 2008; Barnabè et al. 2011; Leier et al. 2016; Nightingale et al. 2019; Sonnenfeld et al. 2019; Shajib et al. 2021).

Strong lensing also provides a critical test of the cold dark matter paradigm by probing the distribution of dark matter in galaxies and galaxy clusters. At large scales, strong lensing constrains the shape and content of dark matter halos (e.g., Kochanek 1991; Koopmans & Treu 2002; Bolton et al. 2006, 2008; Koopmans et al. 2006; Bradač et al. 2008; Grillo et al. 2015; Shu et al. 2016; Shajib et al. 2021). On small scales, strong lensing probes the properties of the small dark matter halos that reside in lens substructure and as field halos along the line of sight (e.g., Vegetti & Koopmans 2009; Vegetti et al. 2010, 2012; Hezaveh et al. 2013, 2016; Birrer et al. 2017; Gilman et al. 2018, 2020; Hsueh et al. 2019; Ritondale et al. 2019b; Meneghetti et al. 2020; Şengül et al. 2022).

---

[70] NHFP Einstein Fellow.

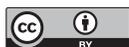
Original content from this work may be used under the terms of the Creative Commons Attribution 4.0 licence. Any further distribution of this work must maintain attribution to the author(s) and the title of the work, journal citation and DOI.

Measurements of strongly lensed sources that are variable in time, such as supernovae and quasars, provide exceptional information about the expansion history of the Universe (Refsdal 1964; Oguri & Marshall 2010; Treu 2010). Time-delay measurements from multiply imaged supernovae (e.g., Goldstein & Nugent 2017; Goldstein et al. 2018; Shu et al. 2018; Pierel & Rodney 2019; Suyu et al. 2020; Huber et al. 2021) and lensed quasars (e.g., Suyu et al. 2010, 2013, 2017; Treu & Marshall 2016; Bonvin et al. 2017; Birrer et al. 2020; Millon et al. 2020; Shajib et al. 2020; Wong et al. 2020) provide independent measurements of the Hubble Constant, $H_0$, that complement measurements from the local distance ladder (e.g., Freedman et al. 2019, 2020; Riess et al. 2019, 2021) and the Cosmic Microwave Background (CMB; e.g., Planck Collaboration et al. 2020). Constraints on $H_0$ from time-delay cosmography using strongly lensed quasars are competitive with other methods (Wong et al. 2020) and are expected to improve as more suitable lens systems are discovered (Shajib et al. 2018), while observations of strongly lensed supernovae may provide a powerful probe in the future (e.g., Suyu et al. 2020). Time-delay cosmography can be used to constrain the expansion history of the Universe and probe the equation of state of dark energy (e.g., Treu 2010; Treu & Marshall 2016; Treu et al. 2018; Birrer & Treu 2021; Sharma & Linder 2022).

The population of candidate strong lens systems has increased rapidly with the advent of deep, wide-field, digital sky surveys. Lens searches with the Sloan Digital Sky Survey (SDSS; e.g., Allam et al. 2007; Estrada et al. 2007; Bolton et al. 2008; Hennawi et al. 2008; Belokurov et al. 2009; Diehl et al. 2009; Kubo et al. 2009, 2010; Lin et al. 2009; Stark et al. 2013; Shu et al. 2017), the CFHTLS Strong Lensing Legacy Survey (e.g., Cabanac et al. 2007; More et al. 2012; Gavazzi et al. 2014; Marshall et al. 2016; More et al. 2016), the Hyper Suprime-Cam Subaru Strategic Program (HSC-SSP; e.g., Sonnenfeld et al. 2018; Jaelani et al. 2020), the Kilo Degree Survey (KiDS; e.g., Petrillo et al. 2017, 2019; Li et al. 2020), Pan-STARRS-1 (PS1; e.g., Berghea et al. 2017; Cañameras et al. 2020), and Gaia (e.g., Lemon et al. 2017; Agnello et al. 2018; Krone-Martins et al. 2018; Delchambre et al. 2019) have yielded thousands of lens candidates and hundreds of confirmed lens systems.

The Dark Energy Camera (DECam; Flaugher et al. 2015) on the 4 m Blanco Telescope at Cerro Tololo Inter-American Observatory in Chile provides one of the premier wide-area imaging systems in the Southern Hemisphere. Searches for strong lenses with DECam have already resulted in thousands of new lens candidates discovered in data from the Dark Energy Survey (DES; e.g., Agnello et al. 2015b; Nord et al. 2016; Diehl et al. 2017; Agnello & Spiniello 2019; Jacobs et al. 2019a, 2019b; O'Donnell et al. 2022; Rojas et al. 2022) and the Dark Energy Camera Legacy Survey (DECaLS; e.g., Huang et al. 2020, 2021; Dawes et al. 2022; Stein et al. 2022; Storfer et al. 2022). However, these searches cover only a fraction of the sky area that DECam has observed, and many more discoveries are expected. Furthermore, searches for strong lens systems with DECam are an excellent precursor for the upcoming Rubin Observatory Legacy Survey of Space and Time (Ivezić et al. 2019), which will cover a similar sky area with much increased sensitivity.

The rapidly increasing quantity and quality of imaging data from current and future surveys continues to provide new opportunities and challenges for strong lens searches. Current ground-based imaging surveys produce catalogs of hundreds of millions of objects, while the relative occurrence rate of strong lensing is $\sim 10^{-5}$ (e.g., Jacobs et al. 2019b). At the same time, the complex morphology of lens systems and the multi-dimensional information content of astronomical imaging (i.e., flux, color, and morphology) makes strong lens searches challenging to automate. Visual searches by groups of experts continue to find hundreds of strong lens candidates (e.g., Diehl et al. 2017; O'Donnell et al. 2022); however, some amount of automation is necessary to fully leverage future large data sets. One attempt to tackle these challenges is by "crowdsourcing" strong lens searches to a large number of human inspectors (e.g., Marshall et al. 2016; Garvin et al. 2022). However, because human visual inspection is subjective, it is often desirable to combine scores in a way that accounts for varying preferences. Intrinsic lens candidate properties may be decoupled from human preferences by using statistical techniques such as matrix factorization (e.g., Mnih & Salakhutdinov 2007).

Another approach to tackle these large data sets is through the adoption of image-based deep-learning algorithms, which have proliferated across astronomy (e.g., Dieleman et al. 2015; Hezaveh et al. 2017; Pasquet et al. 2019; Wu & Boada 2019; Wu et al. 2022). Convolutional neural networks (CNNs; e.g., Lecun et al. 1998; Krizhevsky et al. 2012) in particular have seen much success in recent lens searches (e.g., Agnello et al. 2015a; Bom et al. 2017; Jacobs et al. 2019a, 2019b; Petrillo et al. 2019; Cañameras et al. 2020; Huang et al. 2020, 2021; Li et al. 2020; Dawes et al. 2022; Rojas et al. 2022; Stein et al. 2022; Storfer et al. 2022). The need for automation motivated the community to engage in competitive data challenges to design optimal strong lens-finding algorithms (Metcalf et al. 2019; Bom et al. 2022). Among the lessons learned in the competitions is the fact that human inspection achieved lower performance in huge data sets (Metcalf et al. 2019) compared to CNNs, and that lensing features can be subtle even for visual inspection (Bom et al. 2022). However, when transitioning from simulated data sets to real data, in which there are many orders of magnitudes more nonlenses than lenses, all searches require a final visual inspection for validation (e.g., Huang et al. 2020, 2021; Rojas et al. 2022; Stein et al. 2022), effectively reducing, but not eliminating, the "human in the loop." Thus, it is important to understand both the human and machine biases in lens search procedures.

Here, we present a CNN-based search for gravitational lens systems using DECam data assembled and processed by the DECam Local Volume Exploration Survey (DELVE; Drlica-Wagner et al. 2021). Our search covers a region of the sky that is largely outside of previous strong lens searches using data from DES (DES Collaboration et al. 2021) and DECaLS (Dey et al. 2019), and is largely unexplored by previous deep, ground-based imaging surveys. We specifically target galaxy–galaxy strong lens systems when training our CNN; however, our search also has some sensitivity to lensed quasars and group-scale lenses. Starting from an initial target list of ∼11 million sources, our CNN reduces the candidate list to ∼50,000 candidates. We execute a visual inspection campaign to further refine our list to 581 candidate lens system, which we classify using the ratings assigned during visual inspection. Of these 581 lens candidates, we recover 562 previously undiscovered lenses and 19 previously reported candidates.[71]

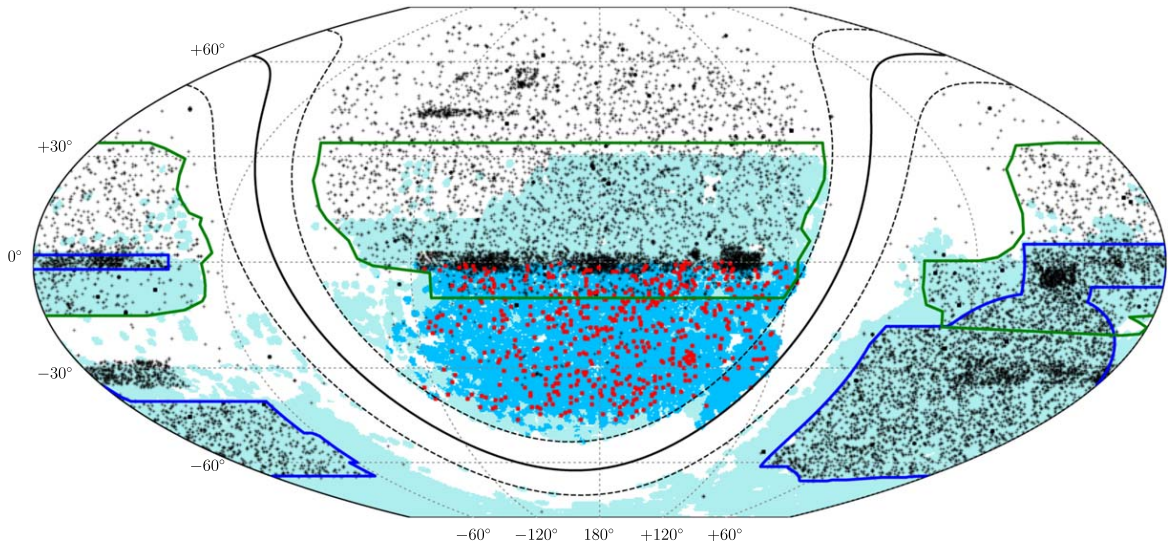

**Figure 1.** Locations of strong lens candidates from this work (red) and previous searches (black) collected in the Master Lens Database (Moustakas et al. 2012) augmented by the results from recent searches using DECaLS (Huang et al. 2020, 2021; Dawes et al. 2022; Storfer et al. 2022), the extended Baryon Oscillation Spectroscopic Survey (eBOSS; Talbot et al. 2021), and HSC (Sonnenfeld et al. 2018, 2020; Chan et al. 2020; Jaelani et al. 2020, 2021; Cañameras et al. 2021; Shu et al. 2022; Wong et al. 2022). Lens candidates are overplotted on the DELVE DR1 *griz* footprint (blue). The DELVE DR2 *griz* footprint (turquoise) is shown for reference, but the search was conducted only on the DR1 data set. The DES footprint is outlined with solid blue lines (DES Collaboration et al. 2021), while the DECaLS region is outlined with solid green lines (Dey et al. 2019). The Galactic plane ($b = 0°$) is shown as a solid black curve, while the two dashed black curves show $b = \pm 10°$. The sky map is shown using an equal-area Thomas–McBryde flat polar quartic projection in celestial equatorial coordinates.

The structure of this paper is as follows. In Section 2, we describe the DELVE data products, our creation of image cutouts, and our pre-processing of those cutouts. In Section 3, we describe the design of our CNN, the generation of our training data, and our training procedure. We present some quantitative evaluations of our CNN performance on both simulated data and previously known lenses. Furthermore, we describe the visual inspection campaign that led to our final ranked candidate list. In Section 4, we describe our sample of 581 candidate strong lens systems and compare to other existing catalogs. Finally we conclude and provide some future outlook in Section 5.

## 2. Data Set

The DELVE data were assembled from multiband imaging by DECam (Flaugher et al. 2015) on the 4 m Blanco Telescope at Cerro Tololo Inter-American Observatory in Chile. DELVE contributes new observations covering a large area of the sky that had not previously been surveyed with DECam. Furthermore, DELVE processes *all* publicly available DECam data with the DES Data Management pipeline (Morganson et al. 2018), thereby providing uniform multifilter imaging of the sky in the *griz* bands to a limiting magnitude of ∼23.5 mag (Drlica-Wagner et al. 2021). The first DELVE data release (DELVE DR1; Drlica-Wagner et al. 2021) provides a catalog of ∼520 million unique astronomical sources assembled from ∼5000 deg² of the high-Galactic-latitude sky in the northern Galactic cap ($b > 10$ and decl. $< 0$). This region is distinct from the DES footprint and overlaps with DECaLS only in the region with decl. $> -7$ (Figure 1).

We begin our search by selecting astronomical objects that conform to a set of color–magnitude cuts developed by Jacobs et al. (2019a) for their strong lens search of the DES data:

$$16 < g < 22$$
$$17.2 < r < 22$$
$$15 < i < 21$$
$$0 < (g - i) < 3$$
$$-0.2 < (g - r) < 1.75.$$

Using a sample of simulated lenses, Jacobs et al. (2019a) estimate that these cuts contain 98.7% of true lenses while significantly reducing the target sample size (and thus the number of false positives). We note that, among the inputs used to create this estimate, the mock deflector galaxies are early-type galaxies with masses drawn from the Hyde & Bernardi (2009) fundamental plane of SDSS elliptical galaxies, and the mock source galaxies are elliptical exponential disks with properties drawn from the Cosmic Evolution Survey (COSMOS) sample (Ilbert et al. 2009). Applying these cuts to the ∼520 million objects in DELVE DR1, in addition to requiring full coverage of the cutout image in each band, *griz*, results in a data set of ∼49 million objects. We apply a cut on the star–galaxy classification (at least one of EXTENDED_CLASS_[GRIZ] ⩾ 2) to select a catalog of extended sources, resulting in a data set of ∼11 million objects. The color–magnitude cuts designed by Jacobs et al. (2019a) were intended to match simulated lens systems around early-type lens galaxies, which are generally massive and have a relatively high lensing cross section. We expect that applying the same color cuts will select a similarly massive set of target galaxies. As a check, we cross-match the galaxies in our target data set against a catalog of stellar mass and photometric redshift for galaxies in DECaLS DR9 (Zou et al. 2022). Figure 2 shows that our target galaxies tend to be more massive and lie at lower redshift relative to the full distribution of galaxies in DECaLS DR9. These conditions indicate a relatively higher lensing cross section, and thus we validate the effectiveness of the color–magnitude cuts.

---
[71] Lens candidates listed in an updated version of the Master Lens Database (Moustakas et al. 2012) augmented with the results from recent searches using DECaLS (Huang et al. 2020, 2021; Dawes et al. 2022; Storfer et al. 2022), eBOSS (Talbot et al. 2021), and HSC (Sonnenfeld et al. 2018, 2020; Chan et al. 2020; Jaelani et al. 2020, 2021; Cañameras et al. 2021; Shu et al. 2022; Wong et al. 2022).

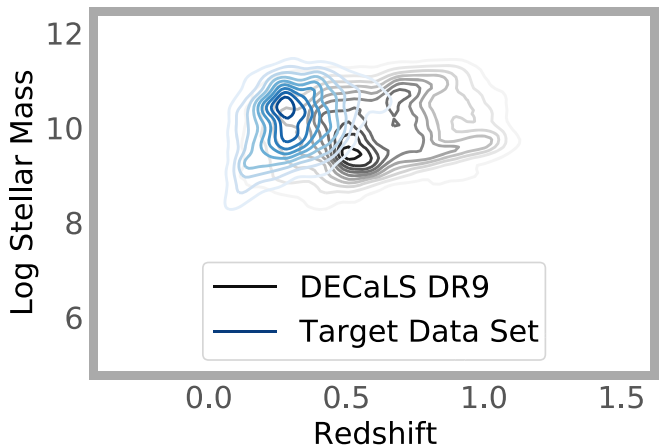

**Figure 2.** Distributions of stellar mass and photometric redshift for both the target data set in this work and the full sample of galaxies in DECaLS DR9, based on the Zou et al. (2022) catalog. Galaxies in our target data set are generally more massive and lie at lower redshift relative to the full distribution of galaxies in DECaLS DR9.

For each object in our initial target list, we determine the best observation of that object in each band. Following the convention defined in Drlica-Wagner et al. (2021), we select the observation with the largest effective exposure time, $t_{\rm eff} \times T_{\rm exp}$, where $t_{\rm eff}$ is the effective exposure timescale factor derived from the full-width at half maximum (FWHM) of the point-spread function (PSF), sky brightness, and extinction due to clouds (Neilsen et al. 2016), and $T_{\rm exp}$ is the shutter-open time of the observation. After identifying the best image in each band, we create $45 \times 45$ pixel ($11\rlap.{''}8 \times 11\rlap.{''}8$) cutout images in each band around each of the sources passing our initial set of photometric selection criteria. We convert the image pixel values, $f$, into flux units, $F$, via the transformation $F = f \times 10^{0.4(30-\mathrm{ZP})}/T_{\rm exp}$ as described in Marshall et al. (2016), where ZP is the zero point derived from the photometric calibration of the image (Drlica-Wagner et al. 2021). It has been shown that rescaling the image data improves network performance (Jacobs et al. 2019b). Therefore, we apply the transformation $X' = (X - \mu)/\sigma$ to each image, where $\mu$ and $\sigma$ are the mean and standard deviation of the image, respectively, and then we rescale each image to the interval [0, 1]. It should be noted that the DELVE DR1 imaging data cover a wide range of exposure times (30–350 s) and PSF image quality (PSF FWHM between $0\rlap.{''}7$ and $2\rlap.{''}0$; Drlica-Wagner et al. 2021). We show the distributions of PSF FWHM and exposure time for our target data set in Appendix A. We found that it was important to capture this variation when constructing the training sample for our CNN-based search described in the next section.

The preceding fully describes the data processing used for CNN training. However, when displaying the images for human visual inspection, we perform the following additional processing. We take our rescaled image data in the *griz* bands and create composite PNG-format images as follows. The channels red ($R$), green ($G$), and blue ($B$) are assigned the bands

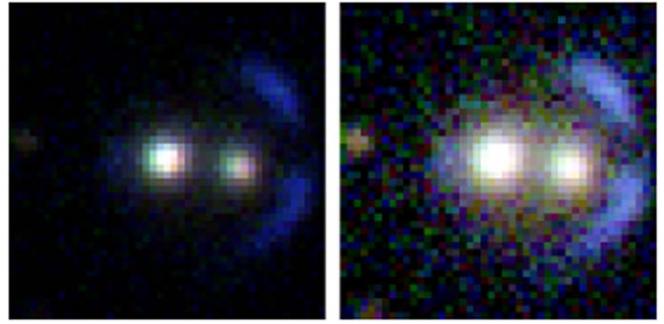

**Figure 3.** Composite PNG image, shown in the two scalings used in this work (Section 2). Left: scaling 1, useful to visualize features near bright objects. Right: scaling 2, useful to visualize very faint features. Both scalings were shown side-by-side during visual inspection of the candidates. Scaling 2 is generally shown in this paper.

$i + z$, $r$, and $g$, respectively. For each channel, we estimate the mean sky value, $\mu$, and sky standard deviation, $\sigma$, using iterative sigma clipping to select mostly sky pixels. Once $\mu$ and $\sigma$ are determined, we first subtract $\mu$ from each channel, then we clip the maximum pixel value at $100\sigma$ and the minimum pixel value at zero, and we finally rescale each channel to integer values in the range 0–255. We find that, to adequately visualize all images, two different "scalings" are useful:

1. Scaling 1: we take the base scaling as described above, and then use the `PIL` package[72] increase the saturation with an enhancement factor of 1.25. This scaling works well to see features in the vicinity of bright objects.
2. Scaling 2: we take the base scaling as described above, and further apply a logarithmic stretch with a stretch factor of 50. We then use the `PIL` package to increase the brightness with an enhancement factor of 1.25. This scaling works well for identifying very faint features.

An example image in both scalings is shown in Figure 3. During visual inspection of the candidates (Section 3.4), we present both scalings side-by-side. Throughout this paper, Scaling 2 is generally used.

### 3. Methods

In this section, we describe our search for strong lens systems in data from DELVE DR1 using a combination of automated and visual inspection techniques. We trained a simple CNN to search for strong lens systems in cutouts created from a sample of ∼11 million astronomical objects. Our CNN was trained on actual DELVE observed images of galaxies from this same data set (negative sample) and the same galaxies with simulated lensing superimposed (positive sample), including training on a false-positive sample (see Section 3.2). We perform visual inspection of the 50,000 highest ranked targets output by the CNN. This visual inspection results in 581 candidates, which are subdivided into three grades. In this section, we describe the construction of our CNN and our visual inspection campaign in more detail.

#### 3.1. Convolutional Neural Networks

We rely on a CNN to perform an initial identification of lens candidates from image data. The core operation in a CNN is a convolution, which essentially measures the level of similarity between a filter (or kernel) and the input image. Thus, CNNs are similar to matched-filter algorithms except that their filters are composed of learnable parameters and can take on any morphological feature. The input can be simultaneously matched to multiple filters, and together these convolution operations form a *convolutional layer*. A nonlinear activation function is applied after each convolutional layer, resulting in a map of positive activations. In order to reduce computation, the activation maps can also be downsampled by using max pooling layers (i.e., keeping only the maximum value in a sliding $2 \times 2$ pixel window). The CNN can be built using a sequence of convolutional layers, activation functions, and pooling layers; after these operations, the activation maps are combined using fully connected layers (i.e., matrix multiplications) followed by activation functions. The final layer outputs two numbers representing the relative likelihood that the input is a lens or a nonlens; by using a softmax function, we can ensure that these likelihoods sum up to unity.

---
[72] http://pil.readthedocs.io

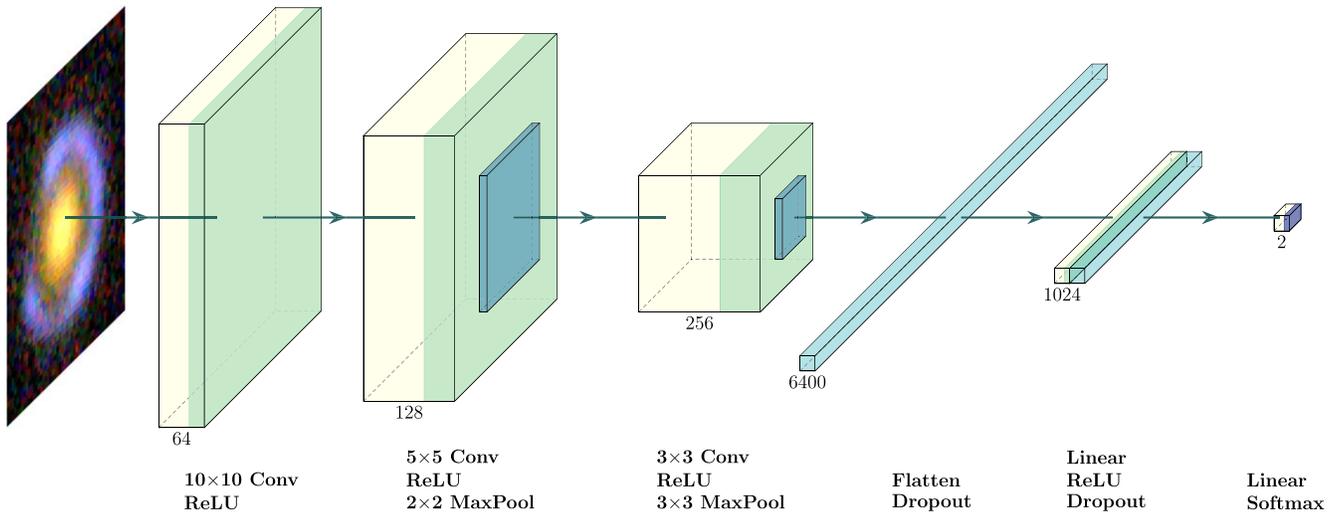

**Figure 4.** Architecture of the CNN used for initial classification of strong lens candidates. The CNN consists of three convolutional layers and two densely connected layers, with dropout layers before each dense layer and max pooling before the latter two convolutional layers. The final dense layer utilizes a softmax activation to normalize the output to the range.

In principle, adding more layers to a neural network (making it "deeper") allows it to learn more complex features and relationships, at the expense of requiring more data to train the additional free parameters. For example, the popular CNN architectures AlexNet (Krizhevsky et al. 2012) and VGG-16 (Simonyan & Zisserman 2014) have eight and sixteen layers, respectively. Some lens searches have used similarly deep networks (e.g., Lanusse et al. 2018); however, other lens searches have shown that relatively simple network architectures with as few as five layers can achieve good performance (e.g., Jacobs et al. 2017).

We opt to use a relatively simple CNN model architecture (Figure 4). Our CNN consists of a total of five layers: three convolutional layers and two densely connected layers, with dropout before each dense layer and max pooling after the latter two convolutional layers. All activation layers used rectified linear unit (ReLU) activations, except for the final dense layer, which used a softmax activation to normalize the output to the range [0, 1].

We conduct supervised learning to train the CNN to robustly identify lens candidates. Training a machine-learning model requires some kind of penalty, or loss function, as well as a method for updating the model parameters based on the penalty. The most common loss function for binary classification as done in this paper is the cross-entropy loss. For our task at hand, the CNN is given examples of lenses and nonlenses and their ground-truth classifications; after it makes its predictions, the CNN incurs penalties for incorrect predictions as well as low-confidence but correct predictions. We use backpropagation and stochastic gradient descent to update the CNN parameters. The backpropagation algorithm computes the contributions to the loss from each of the model's parameters, and stochastic gradient descent determines how each parameter needs to be updated in order to minimize the loss.

We monitor the loss on both the training set as well as a withheld data set used for independently validating performance. After many training epochs (where an epoch is defined as a full pass through the training data set), the training loss will continue to decrease, but the validation loss will eventually plateau and even rise again due to overfitting. A CNN model that is overfit is unlikely to generalize to a novel data set, so we note a few strategies for mitigating overfitting. First, we can use the "early stopping" method to prevent the divergence between training and validation loss where we stop the optimization procedure if the validation loss fails to decrease over a set number of epochs. Second, we can use the "dropout" method, where we randomly drop connections between layers to force the network to come up with redundant combinations of morphological features in order to make predictions. In Section 3.3, we present more details on the CNN training procedure.

### 3.2. Creating the Training Set

The number of known lens systems has reached a sufficient size that some automated lens searches have begun to use images of known lenses for training (e.g., Huang et al. 2020, 2021). However, we opt to use a combination of observed data and simulations in order to include a diverse range of source galaxies, lens galaxies, and lensing configurations. The use of real data captures the inhomogeneous depth and image quality of the DELVE DR1 data set (Section 2). In addition to increasing diversity through sheer quantity, the choice to use simulated lensing mitigates biases and selection effects present in samples of known lenses.

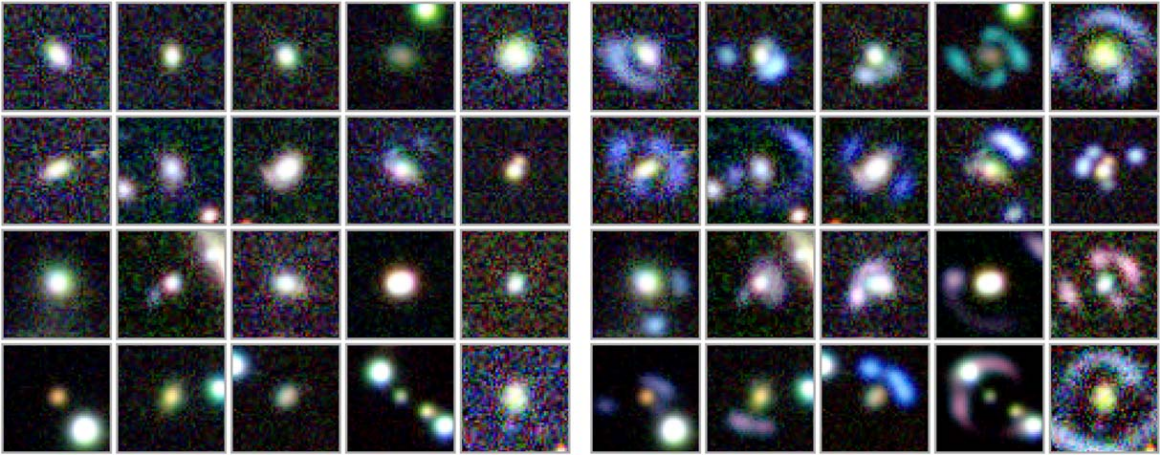

**Figure 5.** Training data examples. Left: negative examples were taken from a random collection of galaxies in DELVE DR1. Right: the same random collection of galaxies with simulated strong lens images generated by `deeplenstronomy`. Cutouts are approximately $11\rlap{.}{''}8 \times 11\rlap{.}{''}8$.

We create our training set by selecting a sample of 20,574 galaxies observed by DELVE DR1 and passing the photometric selection described in Section 2. As discussed in Section 2, we have verified that the galaxies meeting these criteria tend to be relatively massive and therefore are also suitable for use in the training set. We use the unaltered images of these galaxies to serve as negative examples of systems lacking strong lensing. Because of the rarity of strong lensing, very few (if any) true lenses are expected to be found in this sample, and the training should be unaffected by the inclusion of a very small number of false negatives. To create positive training examples of strong lensing, we turn the observed galaxy into a "lens" galaxy by adding simulated lensed sources using the `deeplenstronomy` package (Morgan et al. 2021).

Several measures are taken to enforce realism in the simulated lensing properties. Colors for the simulated sources are drawn from galaxies observed in the DES Y3 Gold Catalog (Sevilla-Noarbe et al. 2021) with DNF_ZMEAN_SOF ⩽ 2.0, where DNF_ZMEAN_SOF is the photometric redshift estimate using the `DNF` algorithm (De Vicente et al. 2016). We additionally filter to DNF_ZMEAN_SOF ⩾ 1.27, which is the minimum source-plane redshift used for strong lensing simulations in Metcalf et al. (2019). Although a neural network would be able to identify very faint lenses, we are also constrained by feasibility of the visual inspection to be performed by humans in order to create a final sample of lens candidates. Thus, we additionally filter on brightness, taking source galaxies with $17.5 \leqslant$ MAG_AUTO_G $\leqslant 22.0$, where MAG_AUTO_G is the apparent magnitude in the $g$ band. The light profiles of the simulated source galaxy are taken to follow a Sérsic profile (Sérsic 1963), parameterized by Sérsic index, $n$, and half-light radius, $R$. We draw $n$ and $R$ from log-normal distributions with $n \sim \ln \mathcal{N}(1.25, 2.25)$ and $R \sim \ln \mathcal{N}(0\rlap{.}{''}5, 2\rlap{.}{''}0)$. These distributions roughly agree with the delensed source properties of lens candidates in similar searches (e.g., Rojas et al. 2022). A comparison between the half-light radii and Sérsic indices of our source galaxies and those of ∼56,000 galaxies from the COSMOS survey that were used for training the `GalSim`[73] software (Rowe et al. 2015) is shown in Appendix A. We assume the magnitude of the external shear, $\gamma$, follows a lognormal distribution with $\gamma \sim \ln \mathcal{N}(0.05, 0.2 \text{ dex})$, which is broadly consistent with the level of external shear expected from ray tracing in N-body simulations (Holder & Schechter 2003; Dalal & Watson 2005). To match the inhomogeneous data characteristics of the DELVE DR1 images, the simulated lensed sources are generated assuming the same exposure time and zero point as the underlying background images, with PSF FWHM sampled from a distribution representative of DELVE DR1.[74] We use a singular isothermal ellipsoid (SIE) parameterization for the lensing potential. For an SIE lensing potential, the deflection of the source light is determined by the Einstein radius, which in turn depends on the redshifts of the source and lens galaxies and the mass of the lens galaxy. Because accurate photometric redshifts and mass estimates are not available for DELVE DR1 galaxies, we instead directly sample the Einstein radius. In order to ensure the simulated lenses display clear lensing features while also fitting within our cutouts, the simulated Einstein radii are constrained to have a log-normal distribution with $R_\mathrm{E} \sim \ln \mathcal{N}(2\rlap{.}{''}5, 1\rlap{.}{''}5)$. As a result, we make no attempt to tie the deflection angles shown in the training images to an estimate of the mass of the lens galaxy. We do not believe this to be an issue as long as the training images are representative of true lens systems with well-separated lensing features. While the model may potentially fail to assign high scores to true lens systems with small source–lens separation angles, we find that those lenses are also more likely to be mistakenly classified as nonlenses by human inspectors during visual inspection. Due to these explicit choices in defining the simulated systems used to train our model, we caution against population-level interpretations of the lens candidate sample produced in this work. In total, we generate 40,000 positive lensing examples and 40,000 negative examples, with each background galaxy image appearing roughly twice in both data sets (though with different lensing features in the positive sample). Examples of our training sample are shown in Figure 5.

While CNNs are adept at learning and identifying the characteristic features of lensing, there are many classes of astronomical objects that can mimic these features and confuse the network, causing it to incorrectly classify them as lenses. Because these objects are far more numerous than strong lenses, it is necessary to harden the network against misclassifying these objects as lenses. First, we train a naive network and use it to score the target sample. The resulting highest-scoring images are dominated by false positives, in particular those that are the most confusing to the model. A sample of these false positives are then used to augment the training set as negative examples. Figure 6 shows the main classes of false positives. They consist of moving objects, dense stellar fields, puffy blue galaxies near red sources, disturbed/merging galaxies, spirals with active star-forming regions, and edge-on blue galaxies near red objects. Eight-hundred and twenty-two false-positive images were flagged for use as negative training examples. In order to increase their relative frequency in the training set, we use the fact that CNNs are not rotation-invariant and create four rotated versions of each false-positive image (0°, 90°, 270°, and 360°) to include a total of 3288 false-positive images. After augmentation, then, we have a total of 43,288 negative training examples.

---

[73] https://github.com/GalSim-developers/GalSim

[74] Generating the simulated sources using the same PSF FWHM as the background images and then re-training gives similar results; roughly 70% of our final candidates and 88% of our Grade A candidates are retained.

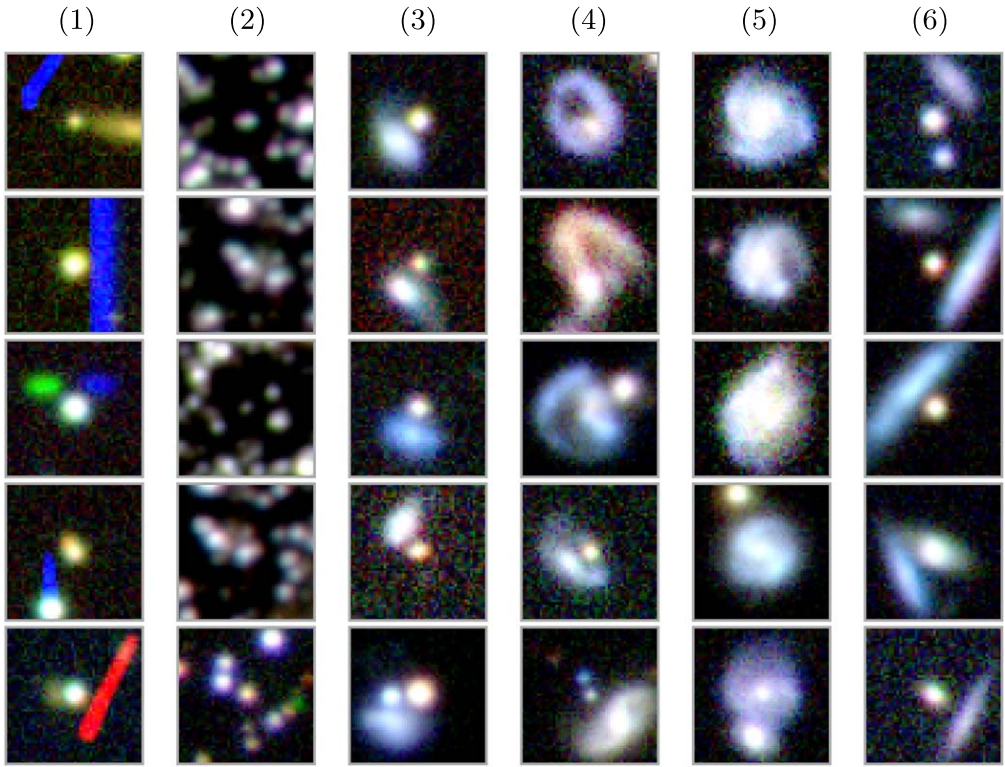

**Figure 6.** Examples of false-positive classes that were used to augment the network training sample. Columns correspond to (1) monocolor streaks from moving objects, (2) crowded stellar fields, (3) puffy blue galaxies near red sources, (4) disturbed/merging galaxies, (5) spirals with active star-forming regions, and (6) edge-on blue galaxies near red objects. Cutouts are approximately $11''\!.8 \times 11''\!.8$.

### 3.3. Training Procedure

We trained our network using 16 Intel Xeon Platinum 8276 2.20 GHz CPU cores on the DES computing cluster at Fermilab. We reserved 25% of the training set as a test sample and further trained the network using a 70%/30% training/validation split, resulting in a net partitioning of 52.5% training/22.5% validation/25% test. We used a batch size of 128 during training. Training was halted when the validation cross-entropy loss failed to decrease by at least 0.0001 for 10 epochs. Our network completed training after 113 epochs. Because the validation loss stopped improving at epoch 103, we used the network at this stage as our final model. This model achieves a validation loss of 0.083, validation accuracy of 0.972, and area under the curve (AUC) of 0.994 using the test sample. Training took approximately 9 hr. Training diagnostics are shown in Figure 7.

As a qualitative check of both overall performance as well as the particular lensing features that our model had learned, we used it to classify a sample of lens candidates identified in DECam data from DECaLS (Huang et al. 2020, 2021). In order to include as many of these candidates as possible for comparison, we took advantage of newer data available in the second DELVE data release (DELVE DR2; Drlica-Wagner et al. 2022), which has larger spatial overlap with DECaLS (Figure 1). Because our search uses a smaller cutout size than the DECaLS searches ($\sim 12''$ versus $\sim 26''$), we visually inspected all of the aforementioned candidates that were available in DELVE DR2 and kept only those that contained visible lensing features in our cutouts. This resulted in a sample of 250 candidates (hereafter the H20+21 sample). An important difference between our search and those outlined in Huang et al. (2020, 2021) is that the latter used a simple magnitude cut $z \leqslant 20$, while we applied a multiband color–magnitude cut (Section 2). Figure 8 illustrates the population sizes after filtering with both sets of cuts, indicating that there is a fairly significant swap population in both cases.

Huang et al. (2020) divide their sample of lens candidates into three grades, summarized below:

1. Grade A—High-confidence candidates. Many of them have one or more prominent arcs, usually blue. The rest have one or more clear arclets, sometimes arranged in

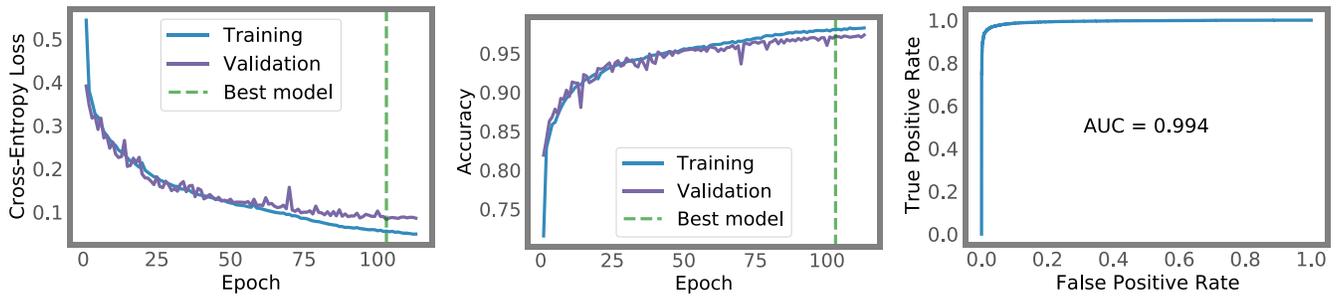

**Figure 7.** Network training and performance diagnostics. Left: cross-entropy loss as a function of training epoch for the training and validation samples. Center: network accuracy as a function of training epoch for training and validation samples. Right: receiver operator characteristic curve for the final trained network using the test sample. The area under the curve (AUC) is 0.994.

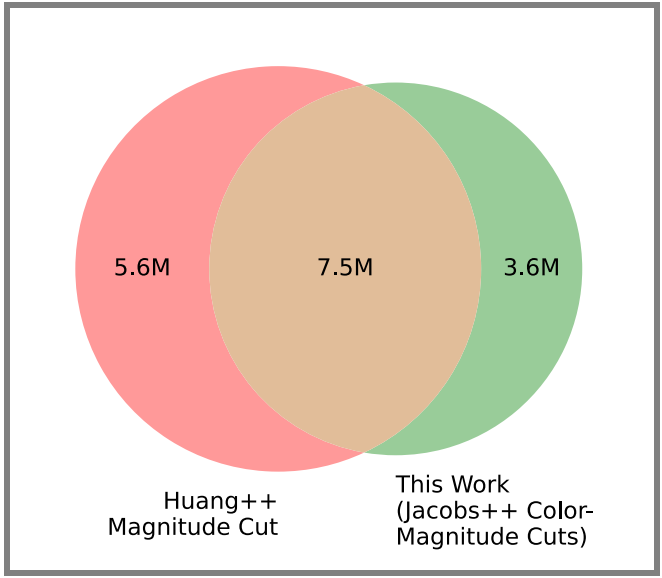

**Figure 8.** Relation between the samples of DELVE DR1 galaxies that pass the $z$-band magnitude cut applied by Huang et al. (2020, 2021) and the galaxy sample that passes our color–magnitude cuts adapted from Jacobs et al. (2019a).

  multiple-image configurations with similar colors (again, typically blue).
  2. Grade B—Characteristics similar to those of the Grade As. For the cutout images where there appear to be giant arcs, they tend to be fainter than those for the Grade As.
  3. Grade C—Generally have features that are fainter and/or smaller than what is typical for Grade B candidates. Counter images are often not present or indiscernible. Putative arclets tend to be smaller and/or fainter, or isolated (without counter images).

We investigate the performance of our network on the H20+21 sample subdivided into these three grades. Figure 9 shows the distributions of scores of the H20+21 sample after scoring with our neural network. For comparison, a red dashed line shows the score cutoff we used to create our candidate list. In total, 128 of the 250 H20+21 candidates (∼51%) pass our score cutoff. Figure 10 shows some of the highest- and lowest-ranking candidates from the H20+21 sample after scoring with our network. Candidates that do not pass the color–magnitude cuts used in this search are outlined in red. Generally, the high-scoring images have large, bright, blue arcs. The low-scoring candidates have a range of morphologies including lensing features with small lens–source separation, lensing features with very large separation (i.e., Einstein radii comparable to the cutout half-length of ∼6″), and lensing features with less blue colors. These images indicate that our model may be less sensitive to the aforementioned features. In Section 4.2, we discuss the outcomes of the H20+21 candidates that appear in our target sample. Aside from a small number of noted exceptions, observing that the neural network generally assigned high scores to images with clear lensing features and low scores to images displaying less-clear or dubious lensing suggested that the network would perform well on the target sample and gave confidence that the training had succeeded. Finally, recalling the variations in depth and image quality in our target sample, we note that we do not observe major biases in the network scores when conditioning on exposure time or seeing (see Appendix A).

### 3.4. Ranking Lens Candidates

Our ranking process and the assembly of our final list of lens candidates involved automated scoring by our CNN and two rounds of visual inspection (Figure 11). To start, we applied the neural network to the target sample of roughly 11 million images. We selected the 50,000 highest-scoring images, which corresponded to a model score cutoff of approximately 0.90. Our score threshold was set dominantly by the volume of lens candidates that we believe could be visually inspected by our team.

The sample of 50,000 highest-scoring images were passed to two rounds of visual inspection (Figure 11). Visual inspection was performed with the `lensrater`[75] tool, with each lens candidate assigned into one of four classes using the following identifications: (0) "not" a lens, (1) "maybe" a lens, (2) "probably" a lens, and (3) "definitely" a lens. A first round of rating consisted of the visual inspection of the 50,000 highest-scoring images, which were divided among eight groups of three human rankers. Included in this set of 50,000 images was a control sample of 2000 images, which were visually inspected by all human rankers. This control sample was intended to allow cross-calibration between human inspectors; however, such comparisons have been left for future work. A total of 5457 images received at least one rating greater than 0; these images were ranked a second time by a uniform group of 12 rankers and again assigned a score of 0 to 3.

To create our final list of lens candidates, we use the final rankings to classify the candidates into grades A, B, and C according to the following criteria:

  1. Grade A
     (a) At least 4 ratings > 2, OR

---
[75] https://github.com/coljac/lensrater

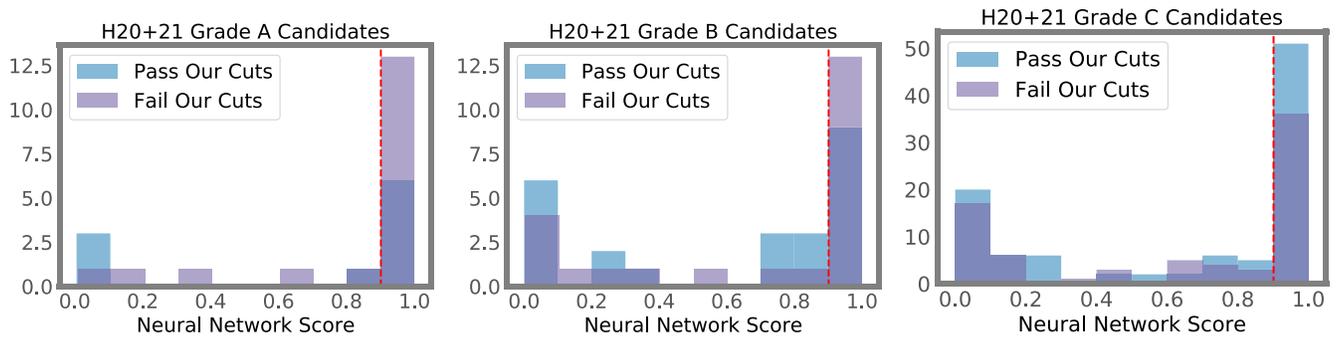

**Figure 9.** Output scores of our CNN classifier applied to Grade A, B, and C lens candidates identified in Huang et al. (2020, 2021). We indicate candidates that pass our color–magnitude preselection (Section 2) in blue and those that fail our selection in purple. The red dashed line indicates the CNN cutoff score that was used to select our candidate list.

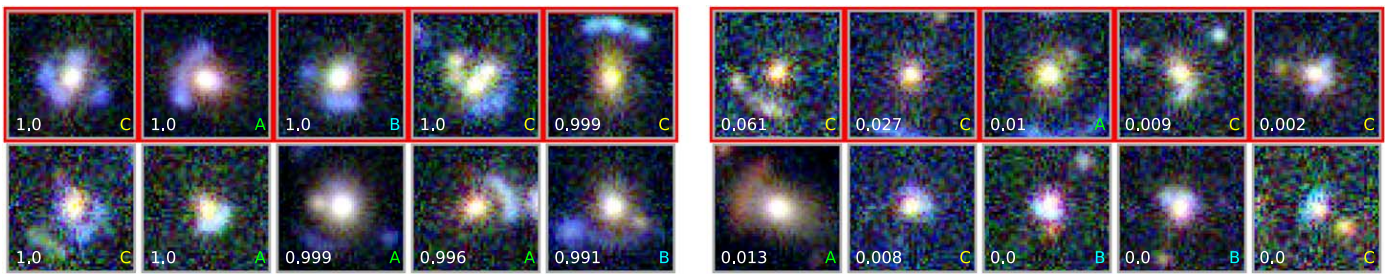

**Figure 10.** Examples of lens candidates from Huang et al. (2020, 2021) that were classified by our network. Images from DELVE DR2, which has a larger overlap with the DECaLS footprint than DELVE DR1 but was not fully searched in this work, are included to gauge model performance. For each image, the score assigned by our network as well as the grade assigned by Huang et al. are shown. Candidates that do not pass the color–magnitude preselection criteria used in this work are outlined in red. Left: Huang candidates that were highly ranked by our network. In general, these candidates show prominent blue arcs that are well separated from the lens. Right: Huang candidates that received a low ranking from our network. These candidates show a combination of close-separation lenses and very far separation lenses, some of which are on the edge of the field of view of our image cutouts. Cutouts are approximately $11\rlap{.}{''}8 \times 11\rlap{.}{''}8$.

    (b) At least 6 ratings $> 1$, OR
    (c) At least 8 ratings $> 0$.
  2. Grade B
    (a) At least 2 ratings $> 2$, OR
    (b) At least 3 ratings $> 1$, OR
    (c) At least 6 ratings $> 0$, OR
    (d) At least 5 ratings $> 0$, of which at least 2 ratings $> 1$.
  3. Grade C
    (a) At least 4 ratings $> 0$, OR
    (b) At least 3 ratings $> 0$, of which at least 2 ratings $> 1$.

For a comparison of the grading criteria described above to a classification by simple mean rating, see Appendix A.

### 3.5. Final Examination of Candidates

Applying the grading scheme in Section 3.4 results in 617 candidate strong lens systems: 60 Grade A candidates, 160 Grade B candidates, and 397 Grade C candidates. As a final check, we create larger cutouts (30″ on a side) for each of these candidates. A total of 536 of the candidates have imaging available in DECaLS DR10,[76] which was released after our search was completed, and in addition to the larger cutout size it also provides deeper coadded imaging compared to the single-epoch imaging in DELVE DR1. For the remaining 81 candidates, we pull larger cutouts from the DELVE DR2 imaging. Three of the authors (F.A., A.D.W., and E.Z.) reviewed the larger cutouts and flagged any candidates that were likely not true lenses based on newly visible features in the larger field of view and/or deeper imaging. This final examination of the candidates resulted in the removal of five Grade A candidates, 11 Grade B candidates, and 20 Grade C candidates from the final list of candidates. Both the original and larger cutouts for each of these removed candidates can be seen in Appendix B.

### 4. Results

Our search yields a catalog of 581 candidate strong lens systems. Using the criteria outlined in Section 3.4, we identify 55 Grade A candidates, 149 Grade B candidates, and 377 Grade C candidates. Additionally, we highlight 8 quadruply lensed quasar candidates in our sample (Section 4.3). A sample of 40 candidates with the highest mean final ratings is shown in Figure 12. The full set of candidates identified in this search is shown in Appendix B. A machine-readable data table containing locations and properties of the candidate lens and source galaxies is available in supplemental material (for a description, see Appendix B). We discuss the properties of our lens candidates in the following sections.

#### 4.1. Properties of Lens Candidates

In Figures 13 and 14, we show the magnitude and color distributions of the lens galaxies in our candidates. We find that the color distribution of lens galaxies appears to track the overall color distribution of galaxies in DELVE DR1. We note that the sharp cutoff in the color distribution of lenses at $(g - r) = 2$ suggests that the true distribution of lenses may continue to redder colors than are allowed by our color preselection. We note that the color cuts that we are using were specified by Jacobs et al. (2019b) on the basis of simulations where the lens–source separation was small. Jacobs et al. (2019b) comment the small lens–source separation may bias the measured colors of their lenses to bluer colors, and we postulate that this may lead to the cutoff that we are observing. Future searches may consider extending the preselection to redder sources, similarly to the approach of Huang et al. (2020).

---

[76] https://www.legacysurvey.org/dr10/description

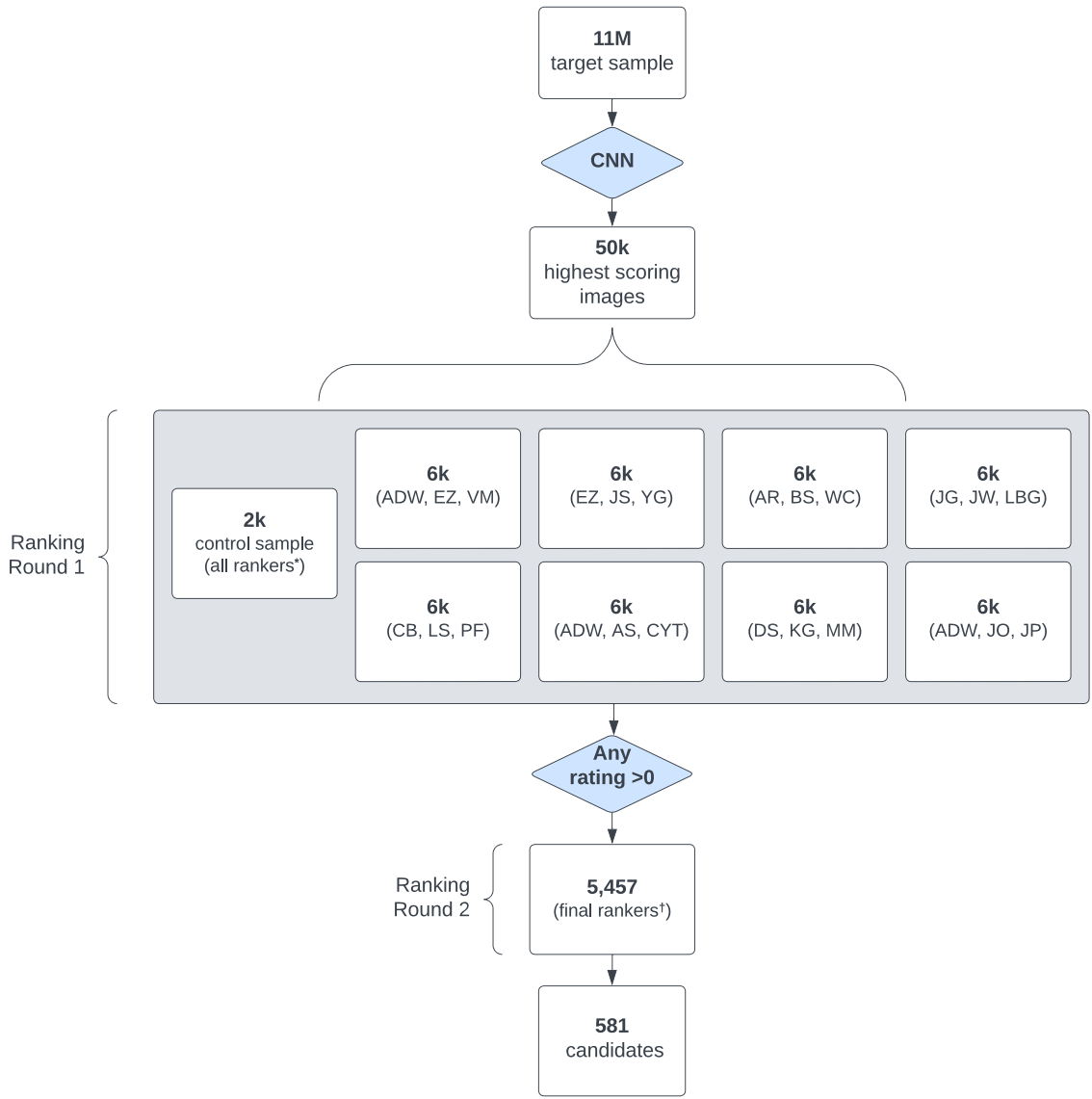

Figure 11. Strong lens candidate selection procedure. Image cutouts were generated for an initial sample of 11 million objects, which were classified by our CNN. The resulting 50,000 highest-scoring candidates were visually inspected and ranked by subsets of the authors. All visual inspectors also scored a control sample of 2000 lens candidates. Out of the 50,000 candidates that were visually inspected, 5457 were assigned a positive classification by at least one inspector. These candidates were passed to a second round of visual inspection, where they were scored by 12 inspectors. This second round of visual inspection resulted in our final set of 581 candidate lenses. Initials of the 24 authors who participated in the visual inspection are shown at the bottom of the figure.

The Einstein radius is a characteristic angle for gravitational lensing and depends on the mass of the lens and the angular diameter distances calculated between the observer, source, and lens. As such, the Einstein radius provides a valuable tool to study cosmological parameters and scaling relationships between galaxy properties and halo mass (e.g., Oguri 2006; More et al. 2016). We follow the procedure of O'Donnell et al. (2022)

and use the average angular separation of source images to approximate the Einstein radius of each lens candidate. Each candidate was inspected manually to identify the positions of the lens and source objects. In the vast majority of cases, the positions of the source images were taken from the corresponding DELVE DR1 catalog entries, found using a query of objects within 6″ separation from the "lens" object the cutout is centered on. Sources that did not appear to be associated with the candidate lens system were removed manually from the Einstein radius calculation. In a small number of cases, a likely lensed source either did not have a catalog detection or had multiple redundant or sporadic catalog detections. In these cases, source detections were added or removed sparingly in order to give a more accurate estimate of the Einstein radius. Finally, in very

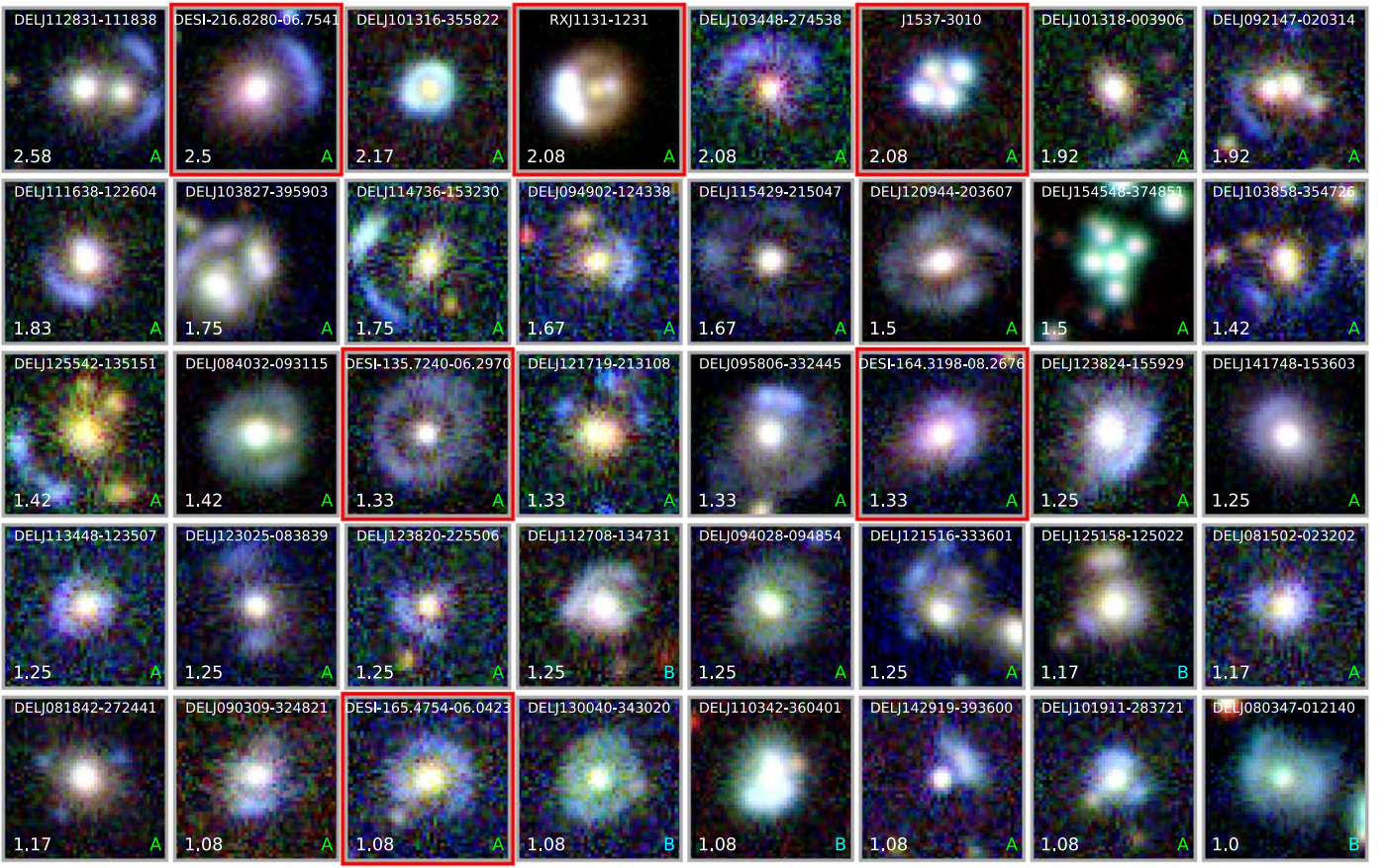

**Figure 12.** Forty candidates identified in this work with the highest mean final ratings. Mean final ratings as well as the grades assigned in this search are shown. Previously discovered lens candidates are outlined in red. Cutouts are approximately $11''\!.8 \times 11''\!.8$.

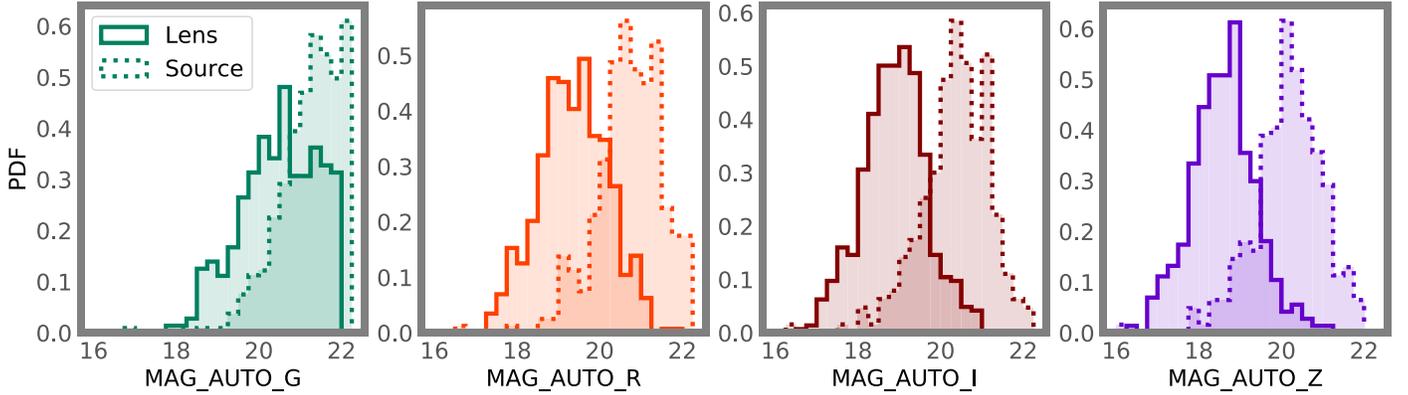

**Figure 13.** Apparent magnitude distributions of DELVE DR1 strong lens candidates in the $g$, $r$, $i$, and $z$ bands (from left to right). Apparent magnitudes are taken from the DELVE DR1 catalog. Distributions of the lens galaxies are shown with solid lines, and distributions of the source galaxies are shown with dotted lines. If a lens candidate has more than one source galaxy, the average apparent magnitude of all associated sources is used.

rare cases, the lens centroid was moved to a different object. For systems with multiple source images, the uncertainty on the Einstein radius is calculated as the standard deviation of the separations. In all cases, the uncertainty is summed in quadrature with the DECam pixel scale ($0''\!.263\,\mathrm{pixel}^{-1}$). For systems with only one lensed image, the uncertainty is taken as 10% of the source–lens separation, which is typical for the scatter of individual sources in systems with multiple images (Shajib et al. 2018; Schmidt et al. 2023). The distribution of estimated Einstein radii is shown in Figure 15. Our estimated Einstein radii are distributed comparably to similar galaxy-scale lens searches (e.g., Rojas et al. 2022). We note that while roughly two-thirds of our candidates have Einstein radius estimates between $1''$ and $3''$, a range that is dominated by massive galaxy-scale lenses, there is a significant tail of candidates with Einstein radius estimates $>3''$ that is likely indicative of group-scale lenses (Cañameras et al. 2020). We also note that the maximum Einstein radius is obviously constrained by the cutout size, and we therefore do not expect our search to recover "giant arcs," which are generally found in cluster-scale lens systems (e.g.,

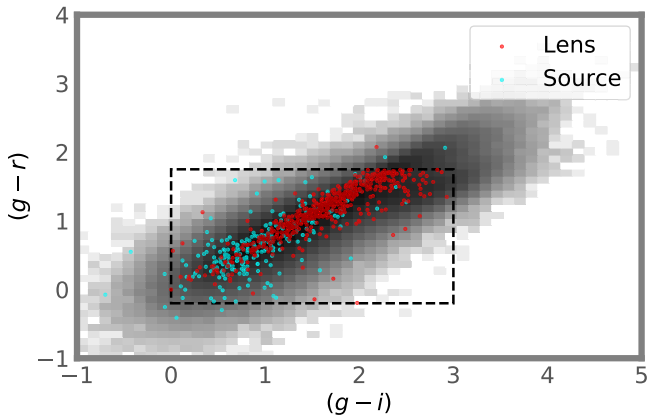

**Figure 14.** Color of DELVE DR1 strong lens candidates. Lens galaxies are indicated by red points and source galaxies are indicated by blue points. The background shows the logarithmic density of a 1% random sample of galaxies from DELVE DR1 and the dashed box indicates the color preselection described in Section 2. If a lens candidate has more than one source galaxy, the average color of all associated sources is used.

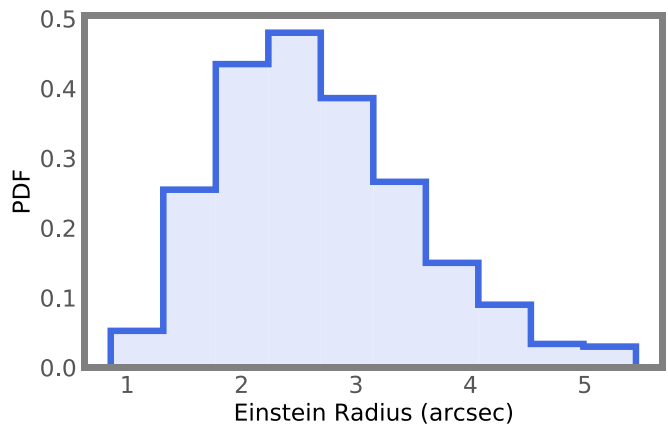

**Figure 15.** Distribution of estimated Einstein radii of the DELVE DR1 strong lens candidates.

Wen et al. 2011; More et al. 2012; Diehl et al. 2017; O'Donnell et al. 2022).

Finally, we draw attention to the fact that candidates displaying apparent Einstein rings may be prone to contamination by ring galaxies. Ring galaxies are a major source of contamination for strong lens searches because of the close similarities in morphology to lensed arcs and Einstein rings. In this work, we attempted to mitigate this to some extent through certain false-positive images shown to the network during training (Section 3.3), and additionally the visual inspectors may have downrated possible ring galaxies in their ratings. We note that there are other approaches to mitigate this source of contamination; for example, Rojas et al. (2022) flag possible ring galaxies and place them into a separate catalog during the visual inspection stage.

### 4.2. Comparison with Other Lens Catalogs

To identify previously discovered lens candidates within our candidate sample, we cross-check against an updated version[77] of the Master Lens database (Moustakas et al. 2012). To ensure that we capture as many previously known lens candidates as possible, we additionally cross-check against several recent lens searches that overlap with the DELVE DR1 area, including those using data from DECaLS (Huang et al. 2020, 2021; Dawes et al. 2022; Storfer et al. 2022), eBOSS (Talbot et al. 2021), and HSC (Sonnenfeld et al. 2018, 2020; Chan et al. 2020; Jaelani et al. 2020, 2021; Cañameras et al. 2021; Shu et al. 2022; Wong et al. 2022). We thus do not expect a significant number of additional matches with candidates from other lens searches. We match 19 of our lens candidates to within 15″ of previously identified candidate systems, with 14 of these systems identified in searches using DECaLS data. The fact that our candidate sample has the highest overlap with searches in DECaLS is perhaps not surprising, due to DECaLS' relatively larger spatial overlap with DELVE DR1, compared to other ground-based surveys, and with the data in this region obtained using the same instrument, DECam.

We therefore compare our final lens candidate sample to the sample of lens candidates identified in similar searches applying neural networks to the DECaLS footprint, which includes H20+H21 (Huang et al. 2020, 2021) and additional candidates from Dawes et al. (2022) and Storfer et al. (2022). This DECaLS sample contains 265 lens candidates located in the region that overlaps DELVE DR1, which we define as $120° < $ R. A. $< 235°$ and $-10° < $ decl. $< 0°$. Of these candidates, 232 are matched within 15″ to objects in DELVE DR1 that have measured fluxes in the $g$, $r$, $i$, and $z$ bands, and 152 of these DECaLS candidates pass our color–magnitude selection (Section 2). However, only 34 of these candidates are scored highly enough by our CNN to be passed to visual inspection, and only 14 of those candidates are included in our final sample. Interestingly, within this same region, our catalog contains 172 lens candidates, meaning that the intersection of our search with previous DECaLS searches corresponds to 8.1% of our candidates in this region and 5.3% of candidates previously identified in DECaLS. We inspect the candidates that were identified by our search but not by previous searches of DECaLS, and we find that they have similar quality to our full sample. Some examples of new candidates identified in the DECaLS region are shown in Figure 16. The presence of high-quality candidates in both samples suggests a high degree of orthogonality between our search and previous searches developed on the DECaLS data.

To better understand the origin of the orthogonality of our search and previous DECaLS searches, we visually inspect the 23 candidates from the H20+H21 sample that passed our preselection criteria and were scored by our CNN. These objects are shown in Figure 17, and they are annotated by the outcome for each step of our ranking, i.e., the score assigned by the CNN, as well as the mean score assigned by the human rankers in both rounds of ranking. Of the 23 candidates from H20+21 in our target sample, 12 passed the CNN score cutoff and were included in the first round of ranking, and eight of these were subsequently assigned at least one rating greater than zero and were included in the second round of ranking. As expected, many images with clear lensing features, generally blue arcs around a yellow or red central object, passed the model score cutoff of ∼0.90 and were subsequently given high scores by the human rankers. Four images passed the initial model score cutoff but were not rated as possible lenses by any of three human rankers in the first round of ranking. These images feature smaller, fainter signs of lensing. The images that did not pass the initial model score cutoff are a mixture of very low-scoring images with lensing features that are difficult to detect, and images that received moderately high scores and do display visible signs of lensing but were not scored highly enough by the CNN to be included in the visual inspection campaign. This latter group may offer some clues about the types of images that were missed by our model (similar to the right-hand side of Figure 10). As noted above, Jacobs et al. (2019a) observed that their color cuts were biased by the lensed source causing additional blueness in the colors of the central object. In Figure 10, several images featuring relatively high-separation arcs are seen to not pass these cuts, which might be explained by the fact that well-separated arcs do not contaminate the colors of the central object to the extent that tighter arcs would. This could also explain why, in Figure 9, it is seen that the H20+21 Grade C candidates, which generally have the smallest lens–source separation, pass these cuts at a higher rate than Grade A or Grade B.

---
[77] Private communication.

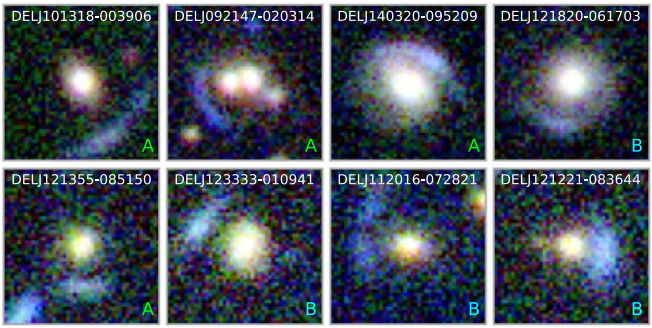

**Figure 16.** Examples of new lens candidates located in the DECaLS footprint that were identified in this work but were not identified in previous searches of that area (Huang et al. 2020, 2021; Dawes et al. 2022; Storfer et al. 2022). The grades assigned in this search are shown. Of the 581 candidates identified in this search, 172 are located in the DECaLS footprint, and only 14 of these were identified in the aforementioned searches.

We note that the orthogonality between our search and previous searches of the DECaLS data comes from a variety of sources. First, the imaging data used for the searches differ, in that previous searches used coadded images in $g$, $r$, and $z$, while our search was performed on single-epoch images in $g$, $r$, $i$, and $z$. There are also differences in distributions of depth and image quality between DELVE DR1 and the DECaLS Data Release 8 (DR8) data used in Huang et al. (2021; see Appendix A). Furthermore, difference in the selection on color, magnitude, and shape of the candidate hosts (Section 3.3) leads to only ∼60% of the DECaLS candidates being passed to our network for scoring. The networks themselves differ both in the network architecture and the data set used for training. For example, Huang et al. (2020) adopt the deeper residual neural network architecture developed by Lanusse et al. (2018), which is a variation of the standard CNN architecture. While our network was trained on a fairly large sample of simulated lenses (Section 3.3), previous searches of DECaLS relied heavily on a smaller sample of real images of confirmed lenses to use as positive examples for training. Finally, once candidates make it to the visual inspection stage, we find that only ∼44% of DECaLS candidates pass our visual inspection, suggesting that there are also differences in the ranking that comes with human inspection. Given the small overlap between candidate samples, we posit that the completeness of future searches may be improved by training networks on a combination of simulated and real lenses and/or applying multiple search strategies to the same sky area.

### 4.3. Quadruply Lensed Quasar Candidates

While our search was specifically tuned for galaxy–galaxy lenses, some of the simulated lens images produced in our training set resemble quadruply lensed quasars (Figure 5). Our human rankers were asked to consider lensed quasar candidates in their ranking, and our network is found to have some sensitivity to known quadruply lensed quasar systems, discussed below. In Figure 18, we show eight candidates identified in the DELVE search that plausibly resemble the morphology of quadruply lensed quasars. We have highlighted these candidates, erring on the side of optimism due to the rarity and scientific value of quadruply lensed quasars.

Based on the TDCOSMO database,[78] two of the candidate systems have been previously identified and subsequently confirmed as lensed quasars. The first is RXJ1131−1231, which was serendipitously discovered during polarimetric imaging of a sample of radio quasars by Sluse et al. (2003) and determined to be a lensed quasar at redshift $z = 0.658$. This lensed quasar system has been the topic of extensive time-delay monitoring and follow-up work in the context of cosmological and astrophysical measurements (e.g., Sluse et al. 2007; Morgan et al. 2008; Chantry et al. 2010; Tewes et al. 2013; Suyu et al. 2017). The second system is J1537−3010, which was independently discovered by Lemon et al. (2019) and Delchambre et al. (2019) using Gaia data. Spectroscopic observations by Lemon et al. (2019) and Stern et al. (2021) concur on a source redshift of $z = 1.72$, and HST observations and automated modeling were performed by Schmidt et al. (2023).

Among our other quadruply lens quasar candidates, we find DELJ154548-374851 and DELJ160650-385722 to be the most promising, showing four point-like sources of similar color tightly clustered around a central object. The other systems show less conventional quadruply lensed quasar morphologies. For example, while DELJ121719-213108 certainly displays four images, it is difficult to tell whether they are point-like, as would be expected for a lensed quasar. DELJ115710-380741 has an unusual configuration if it is a lens; fitting the four image positions using the method developed by Schechter & Wynne (2019), for example, shows significant deviation from a configuration expected for lensing by a singular isothermal elliptical potential. It is possible that this system may actually exhibit two doubly lensed configurations. If DELJ153145-335208 is a true lens, it appears to be very highly distorted, in an extreme shear/ellipticity environment. Finally, the noise level of DELJ085959-325812 makes it difficult to characterize without deeper follow-up imaging.

### 5. Conclusions

We have performed a search for strong gravitational lensing systems in the DELVE DR1 data using a combination of a simple CNN and a campaign of visual inspection. Our CNN was trained on simulated lenses generated by `deeplenstronomy` and superimposed onto the images of real galaxies. We further refined the training by augmenting our training set with examples of some of the most common false-positive artifact classes and validating the network performance on a set of previously known lenses. We applied our trained neural network to ∼11 million cutout images from DELVE DR1 that pass a basic selection in color–magnitude space. The top 50,000 images scored by the CNN were passed through two rounds of visual inspection. The result was the identification of 55 Grade A, 149 Grade B, and 377 Grade C lens candidates, as well as eight candidate quadruply lensed quasars. Due to the location of the DELVE DR1 footprint in the northern Galactic cap with decl. < 0 deg, our candidate list has little overlap with other existing ground-based searches.

---
[78] TDCOSMO collaboration; private communication.

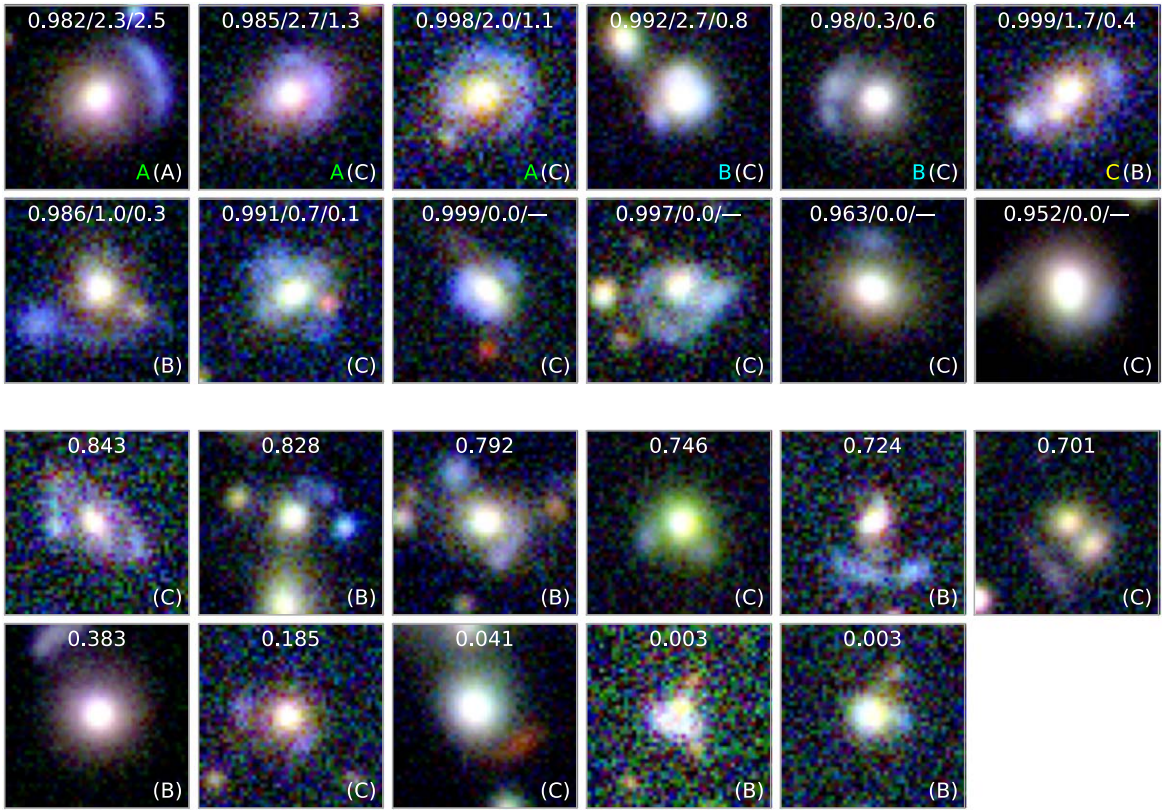

**Figure 17.** Lens candidates from Huang et al. (2020, 2021) that were included in our CNN search of ∼11 million galaxies in DELVE DR1. Top: Huang et al. (2020, 2021) lens candidates that received CNN scores that placed them in the set of 50k images that were visually inspected. Six of these candidates were classified as lens candidates by our visual inspection, while four candidates failed our initial visual inspection. Model score and mean visual inspection scores from the two stages are shown at the top of each image. In each image, the grade assigned by Huang et al. (2020, 2021) is shown in parentheses on the bottom right, and for the six images that appeared in our final candidate list, the grade assigned in this work is additionally displayed with no parentheses. Bottom: Huang et al. (2020, 2021) lens candidates that were not scored within the top 50k candidates by our CNN and were not visually inspected. Model score is shown at the top of each image, and the grade assigned by Huang et al. (2020, 2021) is shown in parentheses on the bottom right. Cutouts are approximately $11''.8 \times 11''.8$.

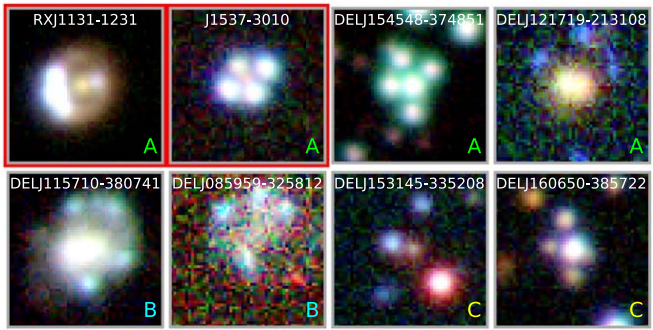

**Figure 18.** Quadruply lensed quasar candidates identified in this work. The first two systems are the previously known lensed quasars RXJ1131−1231 (Sluse et al. 2003) and J1537−3010 (Delchambre et al. 2019; Lemon et al. 2019), while the other six systems have no previous identification. Cutouts are approximately $11''.8 \times 11''.8$.

Looking to the future, DELVE DR2 has already covered ∼17,000 deg² in $g$, $r$, $i$, and $z$. We expect that the final DELVE footprint will cover five times the area of DELVE DR1. The success of our simple CNN suggests that, despite the inhomogeneous depth and data quality of DELVE, it should be possible to recover thousands of galaxy–galaxy strong lens systems by the end of the survey (Collett 2015). The success of strong lens searches in inhomogeneous data from DELVE and other DECam surveys suggest exciting potential for early strong lens searches with the Rubin Observatory LSST before it reaches its nominal depth or uniformity.

## Appendix A
## Supplementary Information

This appendix contains additional figures and information to complement the main text.

As described in Section 2, the DELVE DR1 data set covers a wide range of depth and image quality. The variations in exposure time and seeing are illustrated in Figure A1. Also shown are the same distributions for the DECaLS DR8 imaging data, providing additional context for the comparison of the two data sets in Section 4.2.

To supplement the description of our simulated training set in Section 3.2, here we compare the properties of our simulated source galaxies to those of galaxies found in the COSMOS

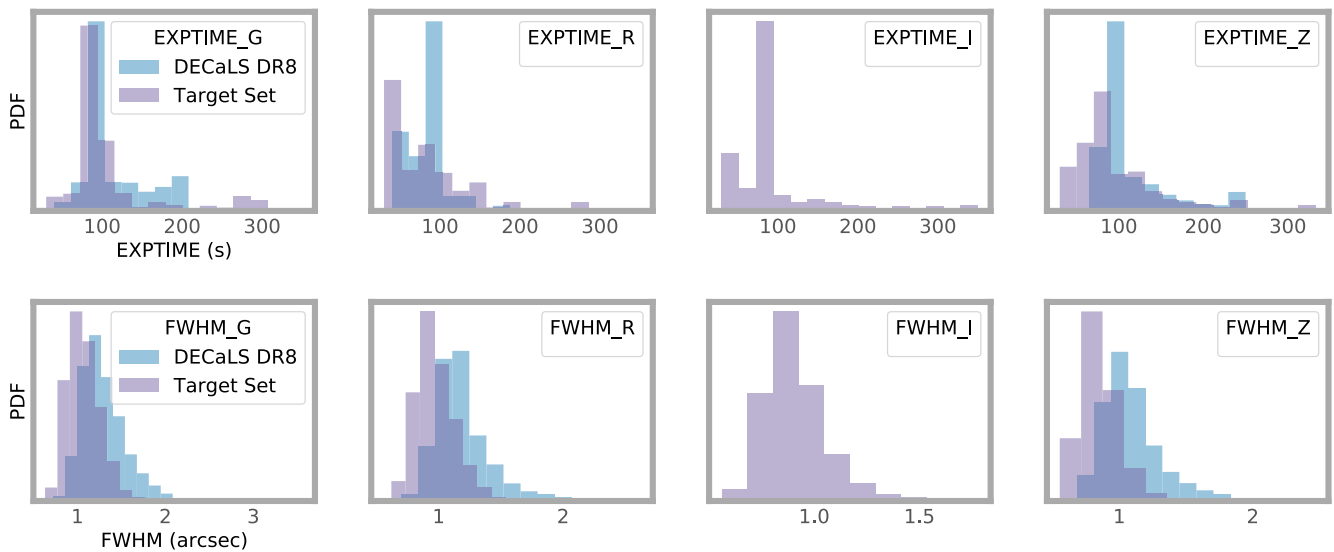

**Figure A1.** Distributions of exposure time (EXPTIME) and seeing (PSF FWHM) in both the target data set from this work and DECaLS DR8, which was the data set used by Huang et al. (2020, 2021). Plots are shown for each of the four photometric bands $g$, $r$, $i$, and $z$.

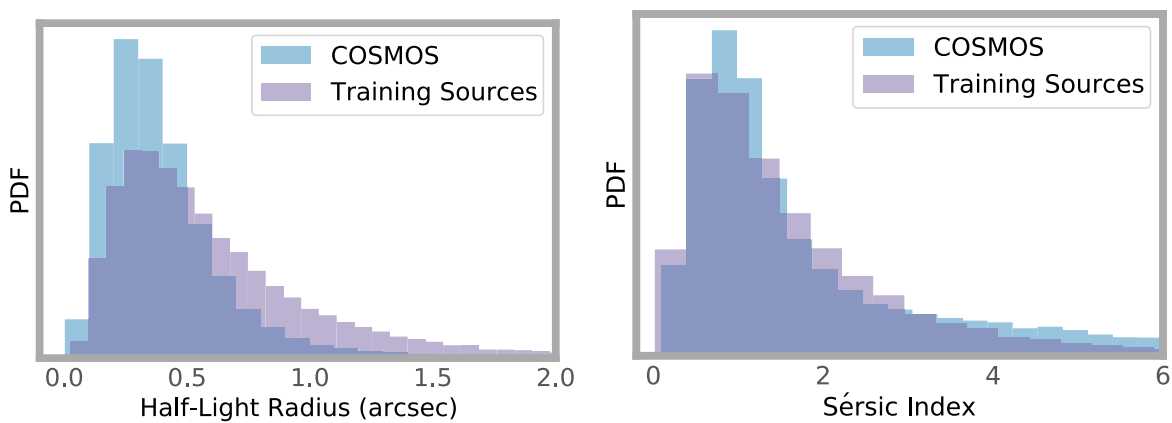

**Figure A2.** Distributions of half-light radius (left) and Sérsic index (right), shown both for the simulated training source galaxies from this work and for ∼56,000 galaxies from the COSMOS survey that were used for developing the GalSim software (Rowe et al. 2015) and training the lens-finding model of Rojas et al. (2022).

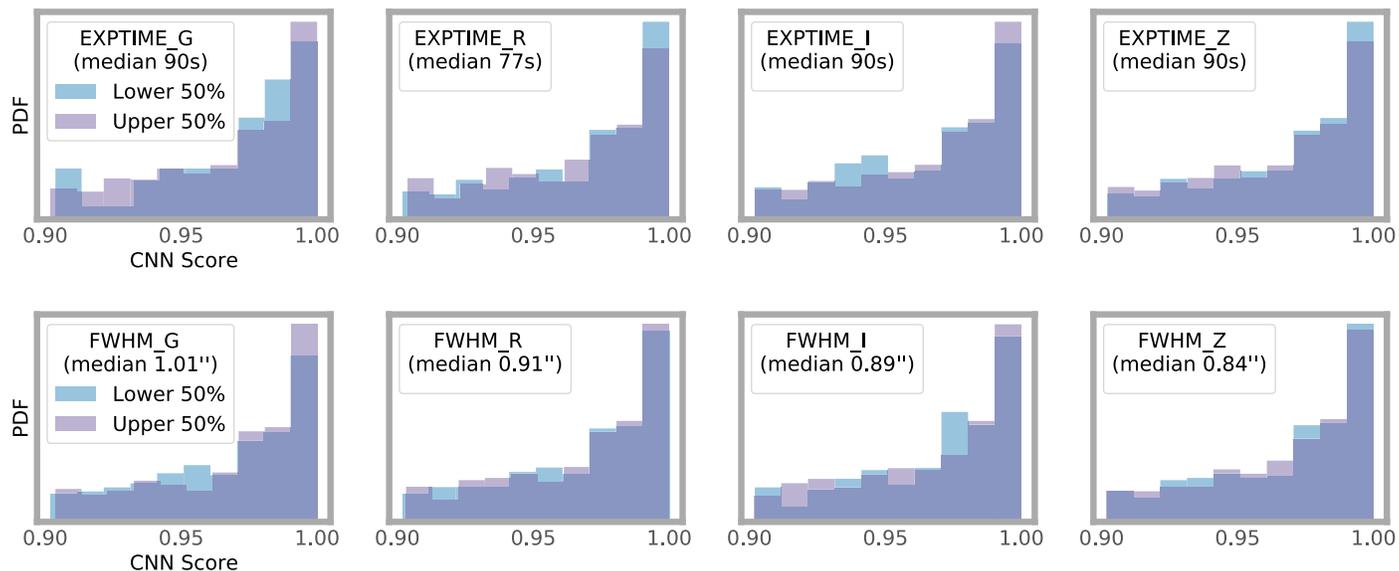

**Figure A3.** Comparison of model scores for images passing our visual inspection score cutoff, conditioned on exposure time (EXPTIME) and seeing (PSF FWHM). Score distributions are overlaid for images constituting the top 50% and bottom 50% of each parameter. Plots are shown for each of the four photometric bands $g$, $r$, $i$, and $z$. No significant biases in the model scores are seen due to these image properties.

Table A1
Comparison of the Grading System Used to Grade Lens Candidates in This Work (Described in Section 3.4), vs. Simple Mean Rating

| Grade (This Work)<br>Grade (Simple Mean) | A | B | C | N/A |
|---|---|---|---|---|
| A | 47 | 13 | | |
| B | 13 | 132 | 16 | |
| C | | 16 | 310 | 73 |
| N/A | | | 73 | 4764 |

**Notes.** Grades assigned according to simple mean rating assume the same number of Grade A, Grade B, and Grade C candidates. The counts of candidates that would be graded the same using either approach are shown in the diagonal entries, while the counts of candidates that would be swapped to different grades are shown in the off-diagonal entries. Here, we have included the 36 candidates that were later flagged as likely nonlenses during a final examination of larger cutouts as described in Section 3.5.

survey. Figure A2 shows comparisons of both half-light radii and Sérsic indices for our simulated source galaxies versus ~56,000 galaxies from the COSMOS survey that were used for developing the GalSim software (Rowe et al. 2015) and training the lens-finding model of Rojas et al. (2022). The half-light radii of the source galaxies used in this work tend toward somewhat larger values than the COSMOS galaxies, although they are roughly consistent with the prior distribution used in Rojas et al. (2022).

As described in Section 3.3, we do not observe a significant dependence of our neural network scores on the inhomogeneous image properties of DELVE DR1. Figure A3 shows distributions of neural network scores for images that passed our visual inspection score cutoff, compared for images above and below the median value of exposure time and PSF FWHM. The distributions are largely similar, illustrating that both subgroups roughly received similar scores from the neural network.

To demonstrate how the grading system used in this work (as described in Section 3.4) compares to an alternative classification by simple mean ranking, Table A1 shows the distribution of grades assigned using both systems. Candidate lens systems that would receive the same grade using either grading system are counted in the diagonal entries, whereas candidates that would receive different grades between the two grading systems are found in the off-diagonal entries.

## Appendix B
## Full List of Lens Candidates

This appendix contains the full set of candidate lens systems identified in this work, as described in Section 4. Table B1 shows a description of the machine-readable data table accompanying this work, which contains locations and properties of our candidates. Finally, we show cutouts of all 581 candidate lens systems. Figure B1 shows 55 Grade A candidates; Figure B2 shows 149 Grade B candidates; Figure B3, Figure B4, and Figure B5 show 377 Grade C candidates; and Figure B6 shows 36 systems that were flagged as likely nonlenses during a final inspection of the candidates, as described in Section 3.5.

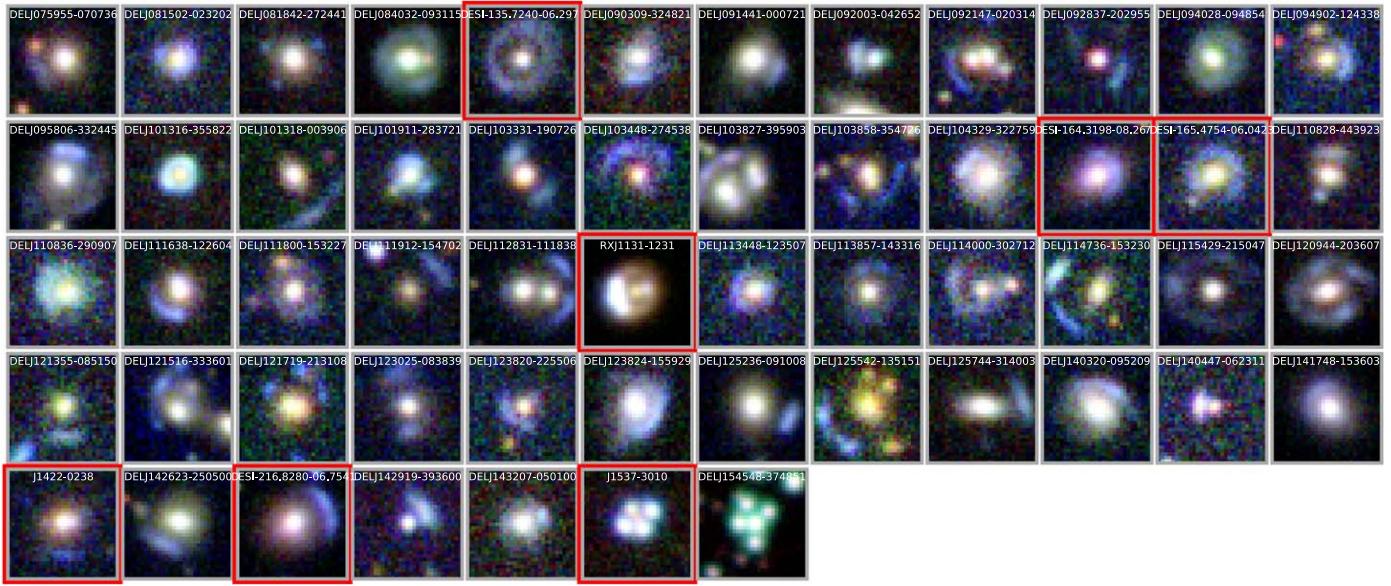

**Figure B1.** Grade A candidates identified in this work. Previously discovered lens candidates are outlined in red. Cutouts are approximately $11''\!.8 \times 11''\!.8$.

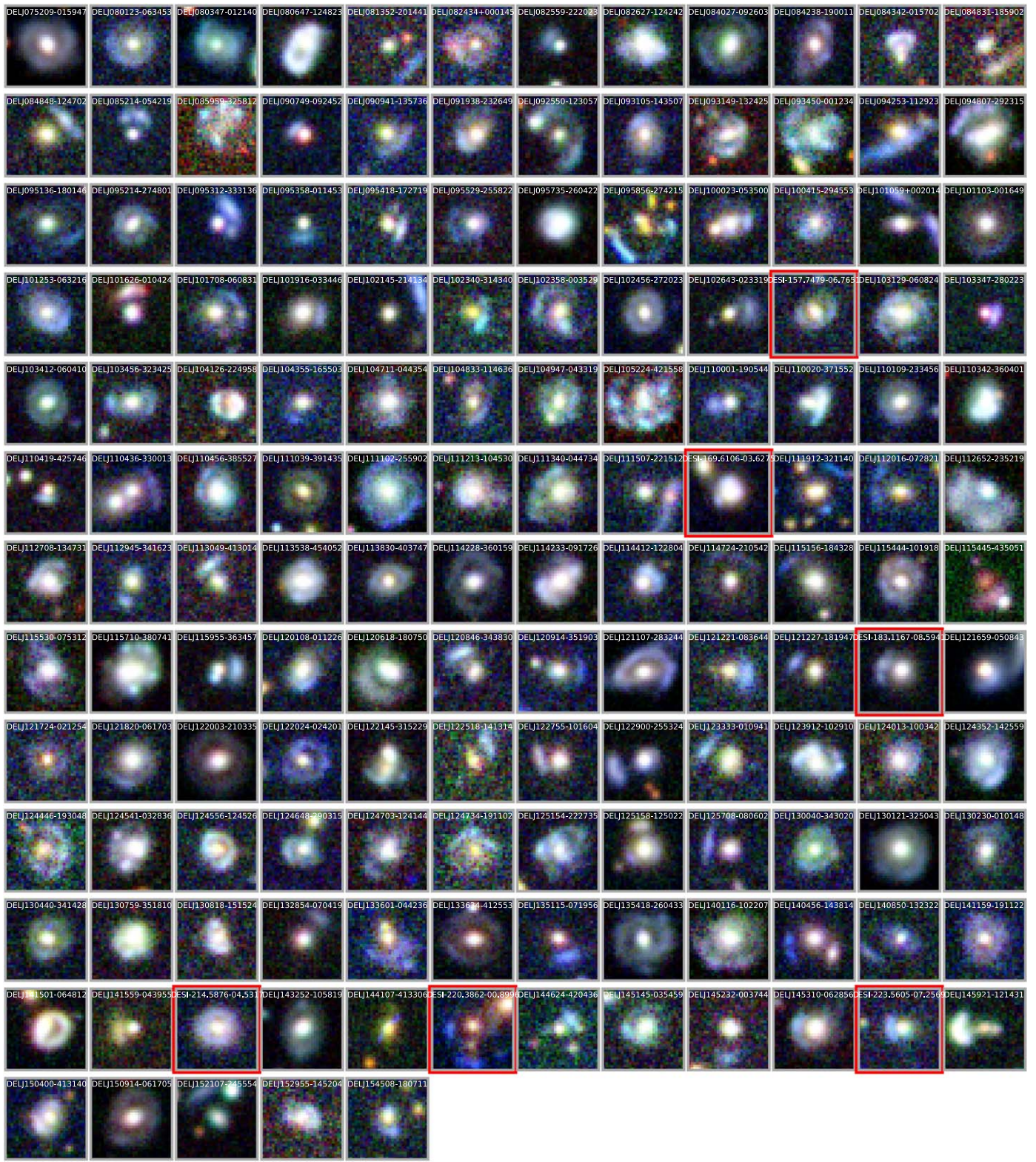

**Figure B2.** Grade B candidates identified in this work. Previously discovered lens candidates are outlined in red. Cutouts are approximately $11''\!.8 \times 11''\!.8$.

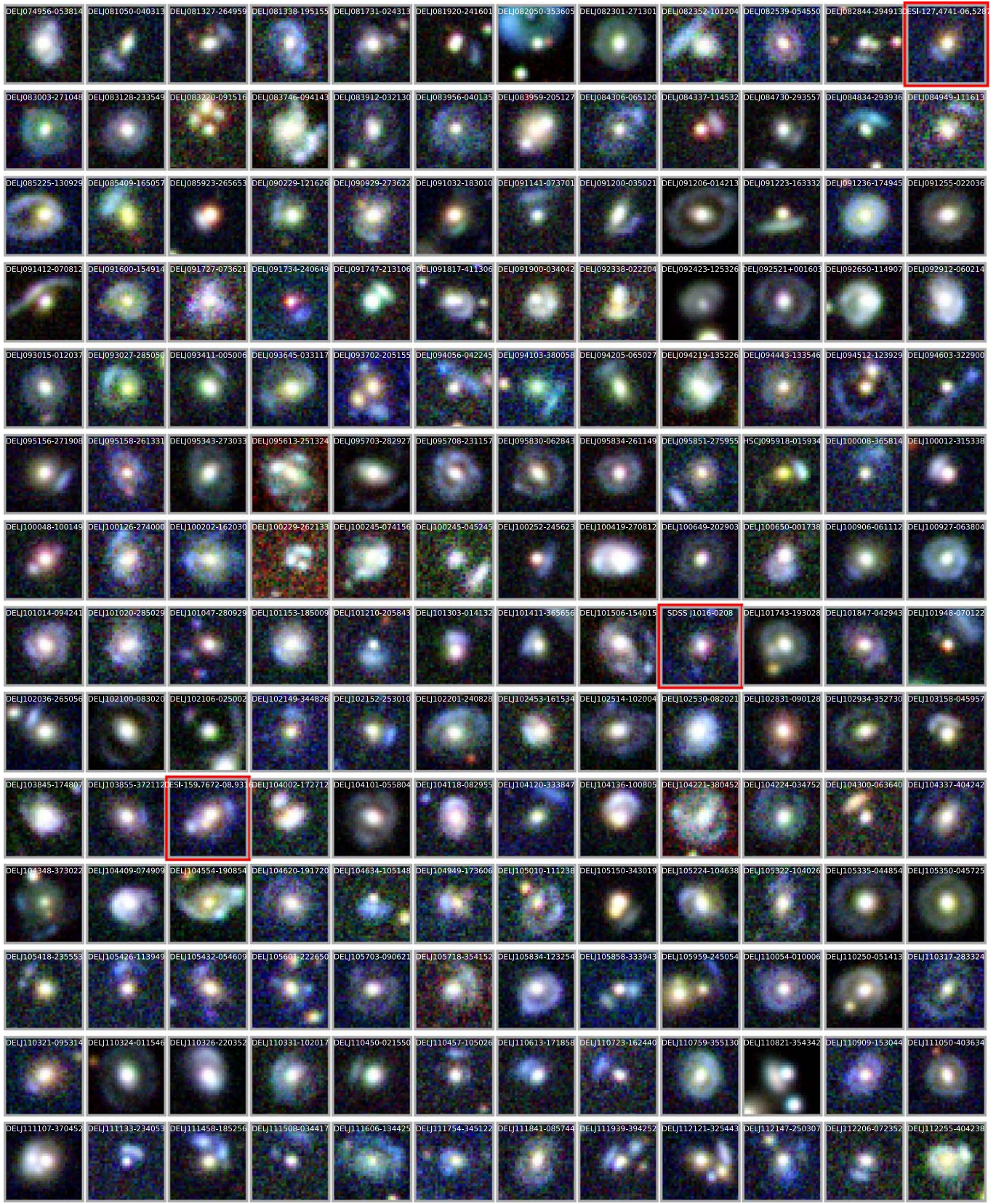

**Figure B3.** Page 1 of grade C candidates identified in this work. Previously discovered lens candidates are outlined in red. Cutouts are approximately $11{''}8 \times 11{''}8$.

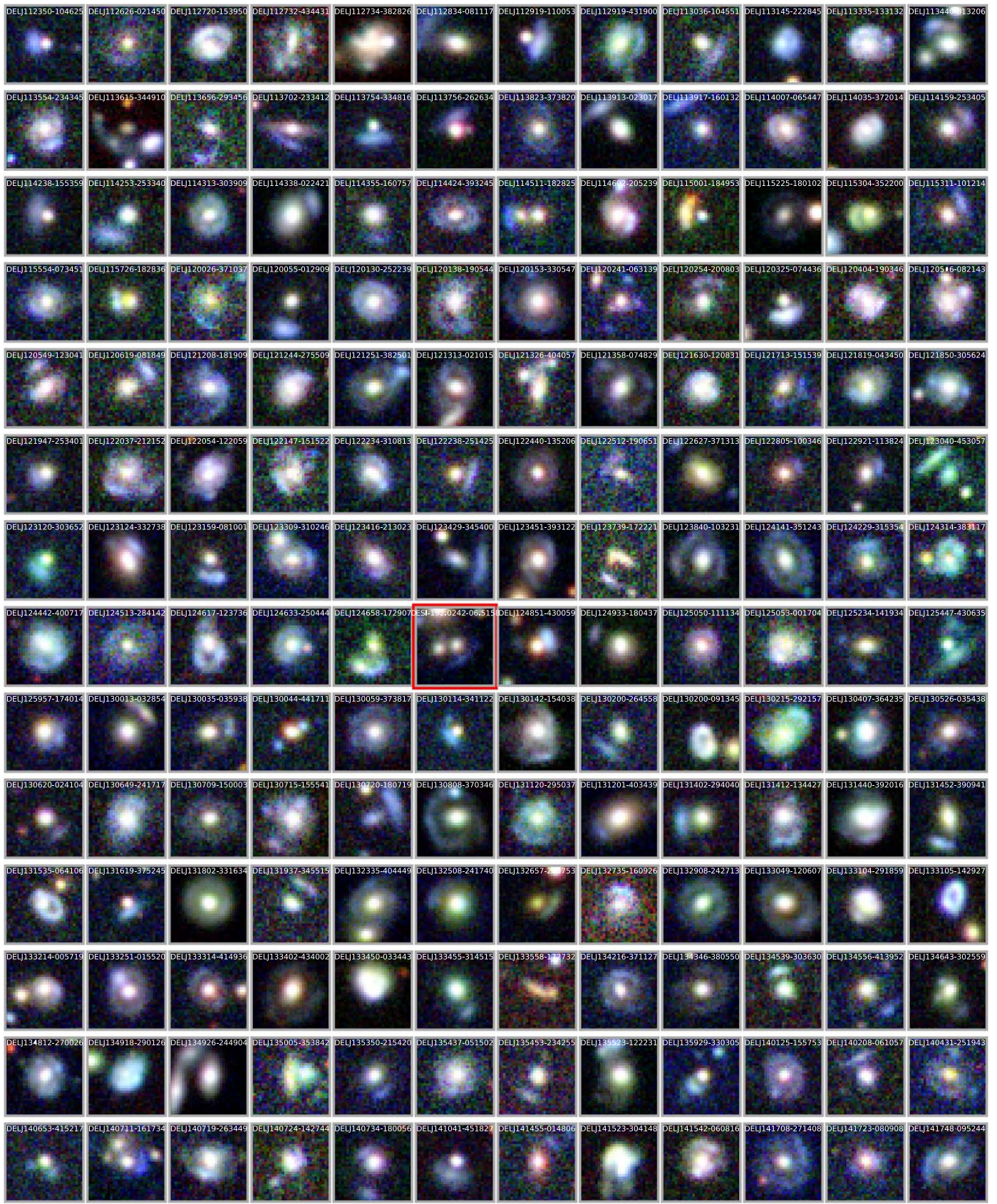

**Figure B4.** Page 2 of grade C candidates identified in this work. Previously discovered lens candidates are outlined in red. Cutouts are approximately $11\rlap{.}''8 \times 11\rlap{.}''8$.

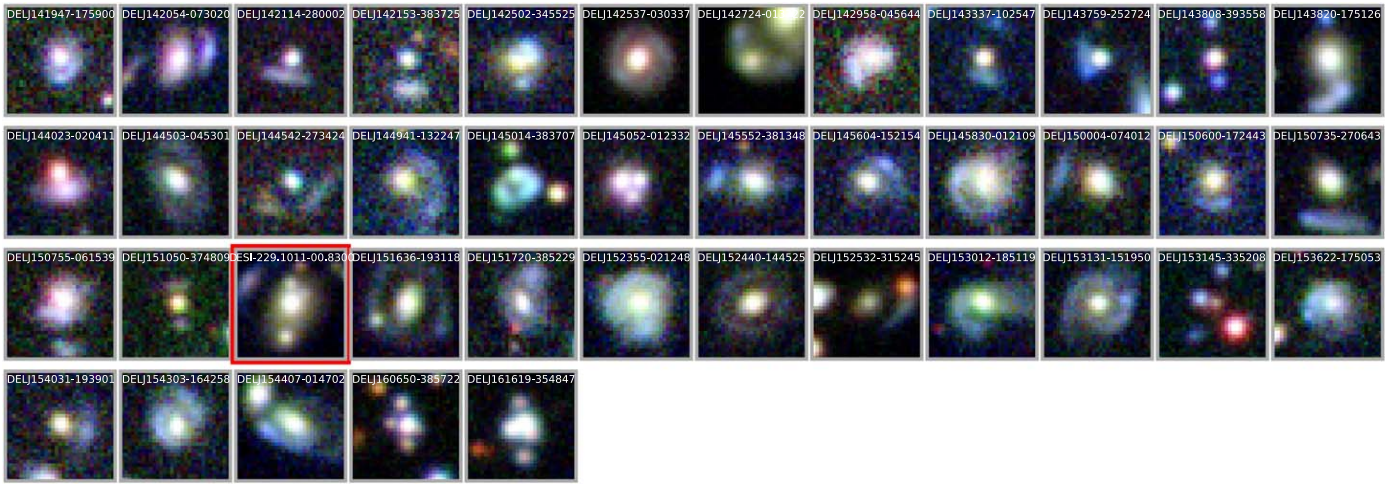

**Figure B5.** Page 3 of grade C candidates identified in this work. Previously discovered lens candidates are outlined in red. Cutouts are approximately $11''\!.8 \times 11''\!.8$.

Table B1
Contents of the Machine-readable Table Available in the Supplemental Materials

| Num | Label | Description |
|---|---|---|
| 1 | name | Lens candidate identifier |
| 2 | mean_final_rating | Mean final rating |
| 3 | grade | Grade (A/B/C/F) |
| 4 | object_type | "lens" or "source" |
| 5 | r_e | Estimated Einstein radius (arcsec) |
| 6 | r_e_err | Uncertainty on estimated Einstein radius (arcsec) |
| 7 | objid_lens | DELVE DR1 Object ID of associated lens galaxy |
| 8 | objid | DELVE DR1 Object ID |
| 9 | ra | R.A. (deg) |
| 10 | dec | decl. (deg) |
| 11 | g | $g$-band magnitude |
| 12 | r | $r$-band magnitude |
| 13 | i | $i$-band magnitude |
| 14 | z | $z$-band magnitude |

**Notes.** The table includes all candidates, including those previously identified in other searches. Candidates that were flagged as likely nonlenses during a final examination of larger cutouts as described in Section 3.5 are included in the table with grade "F." For each candidate system, there are multiple records, one for each associated lens galaxy (object_type = 'lens'') or source galaxy (object_type = 'source'').

(This table is available in its entirety in machine-readable form.)

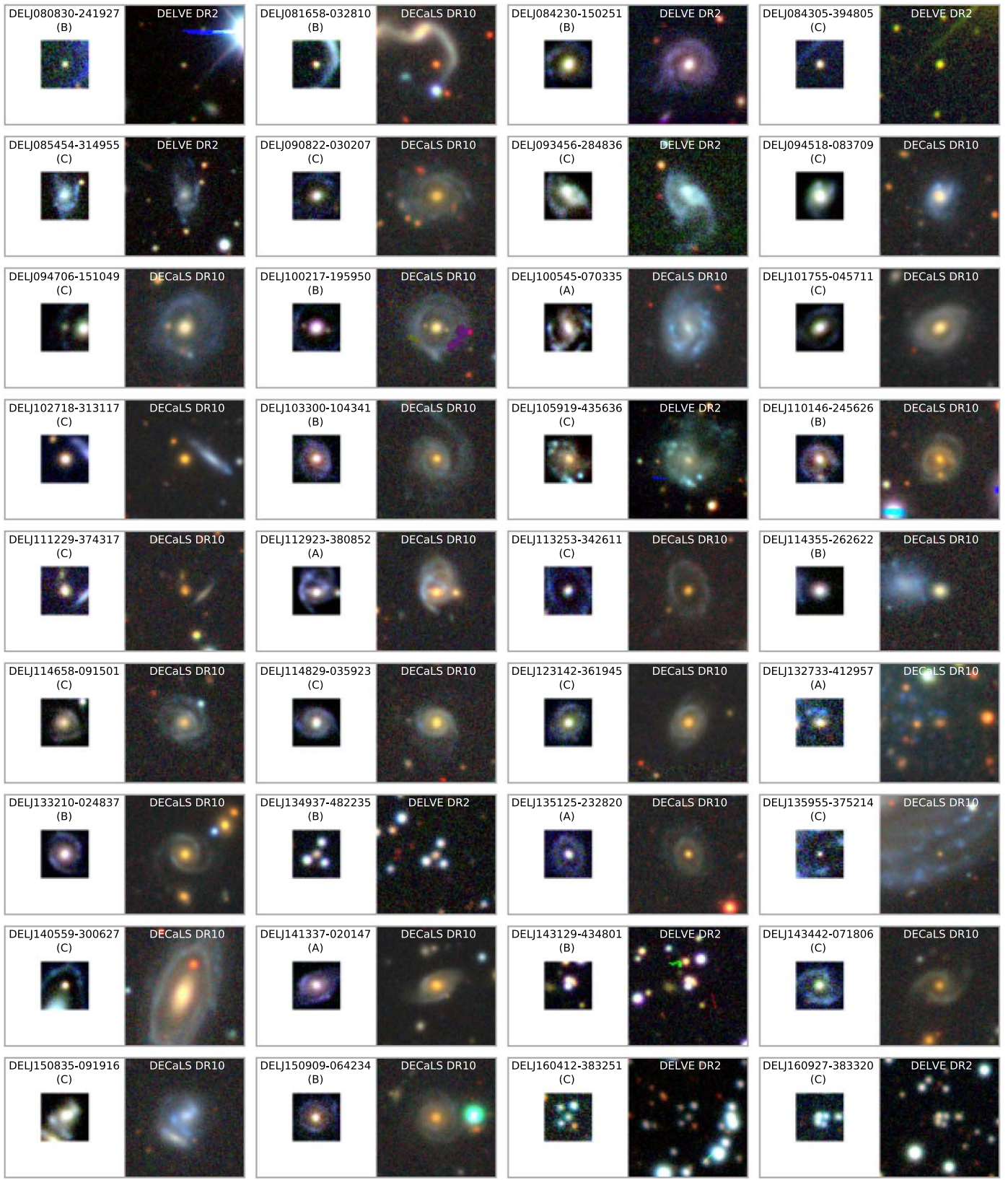

**Figure B6.** Thirty-six systems that were flagged as likely nonlenses based on a final examination of the candidates using larger cutouts as described in Section 3.5. In each box, the original cutout of the candidate and grade assigned are shown on the left. On the right, a larger cutout is shown, along with the data set that it was pulled from. The smaller cutouts are approximately $11\rlap{.}{''}8 \times 11\rlap{.}{''}8$, while the larger cutouts are approximately $30'' \times 30''$.

## ORCID iDs

E. A. Zaborowski https://orcid.org/0000-0002-6779-4277
A. Drlica-Wagner https://orcid.org/0000-0001-8251-933X
J. F. Wu https://orcid.org/0000-0002-5077-881X
R. Morgan https://orcid.org/0000-0002-7016-5471
C. R. Bom https://orcid.org/0000-0003-4383-2969
A. J. Shajib https://orcid.org/0000-0002-5558-888X
S. Birrer https://orcid.org/0000-0003-3195-5507
W. Cerny https://orcid.org/0000-0003-1697-7062
E. J. Buckley-Geer https://orcid.org/0000-0002-3304-0733
B. Mutlu-Pakdil https://orcid.org/0000-0001-9649-4815
P. S. Ferguson https://orcid.org/0000-0001-6957-1627
K. Glazebrook https://orcid.org/0000-0002-3254-9044
S. J. Gonzalez Lozano https://orcid.org/0000-0001-7282-3864
Y. Gordon https://orcid.org/0000-0003-1432-253X
M. Martinez https://orcid.org/0000-0002-8397-8412
J. O'Donnell https://orcid.org/0000-0003-4083-1530
A. Riley https://orcid.org/0000-0001-5805-5766
J. D. Sakowska https://orcid.org/0000-0002-1594-1466
L. Santana-Silva https://orcid.org/0000-0003-3402-6164
D. Sluse https://orcid.org/0000-0001-6116-2095
C. Y. Tan https://orcid.org/0000-0003-0478-0473
E. J. Tollerud https://orcid.org/0000-0002-9599-310X
A. Verma https://orcid.org/0000-0002-0730-0781
J. A. Carballo-Bello https://orcid.org/0000-0002-3690-105X
Y. Choi https://orcid.org/0000-0003-1680-1884
D. J. James https://orcid.org/0000-0001-5160-4486
N. Kuropatkin https://orcid.org/0000-0003-2511-0946
C. E. Martínez-Vázquez https://orcid.org/0000-0002-9144-7726
D. L. Nidever https://orcid.org/0000-0002-1793-3689
N. E. D. Noël https://orcid.org/0000-0002-8282-469X
K. A. G. Olsen https://orcid.org/0000-0002-7134-8296
A. B. Pace https://orcid.org/0000-0002-6021-8760
S. Mau https://orcid.org/0000-0003-3519-4004
B. Yanny https://orcid.org/0000-0002-9541-2678
A. Zenteno https://orcid.org/0000-0001-6455-9135
T. M. C. Abbott https://orcid.org/0000-0003-1587-3931
M. Aguena https://orcid.org/0000-0001-5679-6747
O. Alves https://orcid.org/0000-0002-7394-9466
S. Bocquet https://orcid.org/0000-0002-4900-805X
D. Brooks https://orcid.org/0000-0002-8458-5047
D. L. Burke https://orcid.org/0000-0003-1866-1950
A. Carnero Rosell https://orcid.org/0000-0003-3044-5150
M. Carrasco Kind https://orcid.org/0000-0002-4802-3194
J. Carretero https://orcid.org/0000-0002-3130-0204
F. J. Castander https://orcid.org/0000-0001-7316-4573
C. J. Conselice https://orcid.org/0000-0003-1949-7638
M. Costanzi https://orcid.org/0000-0001-8158-1449
J. De Vicente https://orcid.org/0000-0001-8318-6813
S. Desai https://orcid.org/0000-0002-0466-3288
J. P. Dietrich https://orcid.org/0000-0002-8134-9591
S. Everett https://orcid.org/0000-0002-3745-2882
I. Ferrero https://orcid.org/0000-0002-1295-1132
B. Flaugher https://orcid.org/0000-0002-2367-5049
D. Friedel https://orcid.org/0000-0002-3632-7668
J. Frieman https://orcid.org/0000-0003-4079-3263
J. García-Bellido https://orcid.org/0000-0002-9370-8360
D. Gruen https://orcid.org/0000-0003-3270-7644
R. A. Gruendl https://orcid.org/0000-0002-4588-6517
G. Gutierrez https://orcid.org/0000-0003-0825-0517
S. R. Hinton https://orcid.org/0000-0003-2071-9349
D. L. Hollowood https://orcid.org/0000-0002-9369-4157
K. Honscheid https://orcid.org/0000-0002-6550-2023
K. Kuehn https://orcid.org/0000-0003-0120-0808
H. Lin https://orcid.org/0000-0002-7825-3206
J. L. Marshall https://orcid.org/0000-0003-0710-9474
P. Melchior https://orcid.org/0000-0002-8873-5065
J. Mena-Fernández https://orcid.org/0000-0001-9497-7266
F. Menanteau https://orcid.org/0000-0002-1372-2534
R. Miquel https://orcid.org/0000-0002-6610-4836
A. Palmese https://orcid.org/0000-0002-6011-0530
F. Paz-Chinchón https://orcid.org/0000-0003-1339-2683
A. Pieres https://orcid.org/0000-0001-9186-6042
A. A. Plazas Malagón https://orcid.org/0000-0002-2598-0514
A. K. Romer https://orcid.org/0000-0002-9328-879X
E. Sanchez https://orcid.org/0000-0002-9646-8198
I. Sevilla-Noarbe https://orcid.org/0000-0002-1831-1953
M. Smith https://orcid.org/0000-0002-3321-1432
E. Suchyta https://orcid.org/0000-0002-7047-9358
C. To https://orcid.org/0000-0001-7836-2261
N. Weaverdyck https://orcid.org/0000-0001-9382-5199